\documentclass[osaa,twocolumn,showpacs,8pt,nofootinbib,showkeys,nofleqn]{revtex4-1} 
\usepackage[utf8]{inputenc}
\usepackage[sc]{mathpazo}
\usepackage[T1]{fontenc}
\usepackage{graphicx}  
\usepackage{multirow} 
\usepackage{textcomp} 
\usepackage{dcolumn}
\usepackage{fancyhdr}
\newcolumntype{d}[1]{D{.}{.}{#1}}
\newcommand\T{\rule{0pt}{2.6ex}}       
\newcommand\B{\rule[-1.2ex]{0pt}{0pt}} 
\begin{document} 

\title{On the gas dependence of thermal transpiration and a critical appraisal of correction methods for capacitive diaphragm gauges}

\author{Barth\'el\'emy Daud\'e,$^a$ Hadj Elandaloussi$^{a,b}$ and Christof Janssen$^{a,b}$}
\email[Corresponding author: Christof Janssen, LERMA2, UMR 8112, Tour 32-33 2\textsuperscript{e} ét., 4 pl. Jussieu, 75005 Paris, France]{ christof.janssen@upmc.fr}

\affiliation{$^a$ {\small Sorbonne Universités, UPMC Univ Paris 06, UMR 7092, LPMAA (now LERMA2), Tour 32-33 2\textsuperscript{e} ét., 4 pl. Jussieu, 75005 Paris, France}\\
$^b$ CNRS, UMR 7092, LPMAA (now LERMA2), Tour 32--33 2\textsuperscript{e} ét., 4 pl. Jussieu, 75005  Paris, France}

\begin{abstract}
	Thermal transpiration effects are commonly encountered in low pressure measurements with capacitance diaphragm gauges. They arise from the temperature difference between the measurement volume and the temperature stabilised manometer. Several approaches have been proposed to correct for the pressure difference, but surface and geometric effects usually require that the correction is determined for each gas type and gauge individually. Common (semi) empirical corrections are based on studies of atoms or small molecules. We present a simple calibration method for diaphragm gauges and compare transpiration corrections for argon and styrene at pressures above 1 Pa. We find that characteristic pressures at which the pressure difference reaches half its maximum value, are compatible with the universal scaling $p_{1/2} = 2  \eta\cdot v_{th} / d$, thus essentially depending on gas viscosity $\eta$, thermal molecular speed $v_{th}$ and gauge tubing diameter $d$. This contradicts current recommendations based on the Takaishi and Sensui formula, which show an unphysical scaling with molecular size. Our results support the Miller or \v{S}etina equations where the pressure dependency is basically determined by the Knudsen number. The use of these two schemes is therefore recommended, especially when thermal transpiration has to be predicted for new molecules. Implications for investigations on large polyatomics are discussed.  
\end{abstract}

\maketitle
\thispagestyle{fancy}

\section{Introduction}

Capacitance or capacitive diaphragm gauges (CDGs) are widely utilised pressure sensors for the low to medium vacuum pressure ranges. They combine low relative measurement uncertainty with large dynamic range and high stability. CDG instruments, which are temperature regulated at  above ambient (typically at \( T_2=318.15\)\,K) thus find widespread applications in many areas of metrology and are widely recognised and used as low to medium vacuum transfer standards \cite[e.g.][]{HylandShaffer:1991:Recommended_a,MohanSharma:1996:Comparison_a}.

The measurement principle of these gauges is based on the pressure induced mechanical deflection of an elastic metal or ceramic membrane, which is registered as a change in capacitance of a capacitor of which the membrane constitutes one plate. CDG sensors should thus be highly linear and operate independent of the gas type, but the effect of thermal transpiration, where a temperature gradient creates a pressure difference in a rarified gas,  introduces non-linearity and gas dependence at pressures below about 100\,Pa \cite{Baldwin:1973:Thermal_a,PoulterRodgers:1983:Thermal,Bromberg:1969:Accurate_a}. Thermal transpiration therefore often needs to be accounted for in vapour pressure measurements (e.g. see Refs.~\cite{MartiMauersberger:1993:A-Survey_a,RuzickaFulem:2005:Recommended_a,MokdadGeorgin:2012:Development_a,HansonMauersberger:1985:precision}), or more generally speaking, in investigations of the thermodynamic properties of substances.
Other applications where thermal transpiration has to be considered are accurate scattering and absorption cross section as well as line intensity measurements for atmospheric or other applications -- especially when strongly absorbing species, such as ozone \cite{BarnesMauersberger:1987:temperature} or aromatic compounds are concerned  \cite{FallyCarleer:2009:UV-Fourier_a,ZeccaTrainotti:2012:Positron_a}. This is due to the fact that measurements of these species often require considerable thermal gradients at relatively low pressure. 
But thermal transpiration is not just a phenomenon of metrological interest. Being a special case of non-isothermal rarefied gas flows, thermal transpiration and associated measurements provide additional insight into the larger field of rarefied gas dynamics, which has a wide range of applications in modern vacuum technology and science \cite{Sharipov:1996:Rarefied_a,Sharipov:2012ip,CardenasGraur:2012:An-Experimental_a}. For example, thermal transpiration may allow for new developments for the realisation of thermodynamic motionless micro-machines \cite{PassianWarmack:2003:Thermal_a,McNamaraGianchandani:2005:On-chip_a}.           
           
The phenomenon of thermal transpiration has first been observed and described by Feddersen in 1873 \cite{Feddersen:1873:Uber_a}, but the discovery is generally  attributed to  \citet{Reynolds:1879:On-Certain_a,Reynolds:1879:On-Certain_b} who also coined the terminology.\footnote{See note~2 on page 843 in Ref.~\cite{Reynolds:1879:On-Certain_b}} The phenomenon has then been treated by \citet{Maxwell:1879:On-Stresses_a} and  \citet{Knudsen:1909:Eine_a} and there have been a large number of experimental, analytical and numerical investigations since. Numerical calculations have reached a level of sophistication which can show high degree of agreement between theory and experiments \cite{VaroutisValougeorgis:2009:Simulation_a,CardenasGraur:2012:An-Experimental_a}, but the treatment of polyatomic gases poses fundamental difficulties and even for diatomic molecules it is an open question whether the degree of agreement with experiments can exceed several tens of percent \cite{LoyalkaStorvick:1982:Thermal_a,TitarevShakhov:2012:Poiseuille_a}. Moreover, quantitative predictions based on  numerical approaches  either are tedious or they still depend on experimentally determined parameters which, in turn, are determined from thermal transpiration measurements and which, once more, have only been verified on atoms and relatively small molecules \cite{Sharipov:2011:Data_a,SharipovSeleznev:1998:Data_a}. Therefore, the most common corrections to apply to CDG measurements are based on semi-empirical approaches \cite{PoulterRodgers:1983:Thermal,JitschinRohl:1987:Quantitative,YoshidaKomatsu:2010:Compensation_a}, of which the equation due to Takaishi and Sensui (or TS hereafter) \cite{TakaishiSensui:1963:Thermal} is the most frequently used -- even though some critics have been raised recently \cite{CardenasGraur:2012:An-Experimental_a,Setina:1999:New-approach_a}. The TS approach, a recent modification by \v{S}etina \cite{Setina:1999:New-approach_a} as well as the formula of Miller \cite{Miller:1963:On-the-Calculation_a}  have been shown to be particularly adapted for pressure corrections using nitrogen as the measurement gas \cite{YoshidaKomatsu:2010:Compensation_a}.

The different schemes express the pressure ratio 
\begin{equation}   \label{eq:DefPRat} 
	R=p_1/p_2
\end{equation}  between two volume elements at two different temperatures \( T_1 < T_2 \) and connected by a tube of diameter \( d \) as a function of the pressure in the sensor \( p_2 \) over a pressure range that varies from viscous to molecular flow regimes. By convention \cite[e.g.][]{TakaishiSensui:1963:Thermal,Miller:1963:On-the-Calculation_a}, $p_2$ is the pressure that is directly accessible by measurement (and $R = p_2/p_1$ if $T_1 > T_2$). The low pressure limiting value \( R_0 \) may reach the Knudsen ratio 
\begin{equation}    \label{eq:KnudsenRat}
R_{\mathrm{K}} = \sqrt{T_1/T_2} \leq R_0,
\end{equation} which amounts to a pressure correction of up to 3.5\,\% under typical laboratory conditions (\( T_1 = 296.15\,\mathrm{K},\ T_2 = 318.15\,\mathrm{K} \)) for pressure measurements using CDGs, but deviations from the low pressure limiting Knudsen ratio due to the neglect of the details of the molecule surface interactions have been demonstrated both theoretically \cite{Siu:1973:Equations_a,NishizawaHirata:2002:DSMC_a} and experimentally \cite{Hobson:1969:Surface_a}.
Another common simplification is the use of a single characteristic diameter \( d \) instead of taking into account the exact geometry which might be much more complex. This has led to identifying \( d \) as an effective parameter rather than the geometric dimension of the narrowest element \cite{JitschinRohl:1987:Quantitative}.  
                                                      
A limitation for the direct application of available correction schemes is that these have been tested with only a few and mostly small, \emph{ie} rare gas or diatomic molecules. The equation proposed by \v{S}etina \cite{Setina:1999:New-approach_a}, for instance, has so far been verified on just the four gases Ar, H\( _2 \), He and N\( _2 \). Other approaches have been tested on some more and also larger molecules: the CH\( _4\) molecule, for example, has been investigated repeatedly \cite{TakaishiSensui:1963:Thermal,Furuyama:1977:Measurements_a,Yasumoto:1980:Thermal_a,TejucaPajares:1976:Thermal_a}, as well as SF\( _6 \) and  C\( _2 \)H\( _6 \) \cite{PoulterRodgers:1983:Thermal,TejucaPajares:1976:Thermal_a}. The most extensive study of the TS equation in terms of number of molecules has been performed by Yasumoto \cite{Yasumoto:1980:Thermal_a}. In his study 23 condensible and non-condensible molecules including several non-methane hydrocarbons with up to 14 atoms (butane) were employed. Still, the results are somewhat contradictory inasmuch as they show a much weaker pressure dependence than the original measurements of Takaishi and Sensui \cite{TakaishiSensui:1963:Thermal} or those of Yoshida \emph{et al.} \cite{YoshidaKomatsu:2010:Compensation_a}.  Moreover, the derived dependence on the molecular diameter is only partly consistent with the experimental observations. Finally, unlike many other approaches, neither the parameterisation proposed by Yasumoto \cite{Yasumoto:1980:Thermal_a} nor the original TS parameterisation can be cast in a form that depends exclusively on the Knudsen number $Kn$ (ratio of mean free path over diameter \(Kn =\lambda / d \)), which is difficult to conceive theoretically. 
Such a $Kn$ dependence would indeed be expected, because thermal transpiration is caused by a temperature gradient driven creep flow. This flow creates a  pressure gradient which maintains a counterbalancing mass motion. 
Another concern with the original proposition of Takaishi and Sensui is that it seems to break down for large molecules, where one of the parameters changes sign (see section~\ref{sec:CorrectionSchemes}). 

In the light of the fact that the TS correction is generally recommended and most widely adopted, and recognising that on the one hand other approaches have rarely been tested on organic molecules but that on the other hand corrections for larger molecules have become increasingly important \cite{ZeccaChiari:2011:Total_a,ZeccaTrainotti:2012:Positron_a,FallyCarleer:2009:UV-Fourier_a}, it seems to be just timely to verify the validity of the three above correction schemes to larger molecules (with molecular diameter \( D \simeq 500 \)\,pm or more). We are not aware that such a comparison has been attempted before. Earlier studies either compared  different approaches using much smaller (diatomic) molecules or investigated the gas dependence using only a single approach.

In this article, we thus study the gas dependence of thermal transpiration equations by measuring the transpiration effect in the 1$-$130\,Pa pressure range using the two gases argon (Ar) and styrene (C\( _8 \)H\( _8 \)). We first present a short overview of proposed correction equations  and discuss their gas dependencies based on the pressures at which thermal transpiration becomes important. We then describe our measurements and confront the results with the different schemes. Our measurements rely on the comparison of two CDGs, one of them being operated at ambient temperature without  stabilisation and thus requiring in-situ calibration. The new calibration method, which can be easily put into place, is verified by comparing thermal transpiration measurements of Ar with the numerous results available in the literature. The results on styrene will be used for an appraisal of the three most common correction schemes and for identifying those who apply best to the experimental situation.                                                                      
                                                                        
 \section{Empirical treatments of thermal transpiration}  \label{sec:CorrectionSchemes}
   
 A wealth of empirical and semi-empirical formulas have been proposed to describe the thermal transpiration effect.  Here, or in the Appendix, we will give a short account of these, because the single detailed overview by \citet{YoshidaKomatsu:2010:Compensation_a} is only available in the Japanese language. 
Several of the transpiration equations arise as approximate solutions of the following differential equation
 \begin{equation}  
	\label{eq:diffthermaltrans} \frac{d p}{p} = \Theta( d /\lambda ) \frac{1}{2}\frac{ d T}{T} ,
\end{equation}
with suitable  \( \Theta(d/\lambda) \) and where \( d \) and \( \lambda \) denote tube diameter and mean free path, respectively.   \(  \Theta( d /\lambda )   \) is an inverted-S shaped transition function which must take the limits 1 and 0, for \( d /\lambda  \ll 1 \)  and  \(\gg 1 \), respectively, corresponding to the values \( R_K \) and 1 for the pressure ratio \( R \). 
Knudsen \cite{Knudsen:1909:Eine_a} derived the above expression with \( \Theta(d/\lambda)= \left(1+d/\lambda\right)^{-1} \) for cylindrical tubes at low pressures \( (d < \lambda) \). Generally, \( \Theta \) depends on pressure and temperature, which complicates finding closed analytic expressions and many different approximations have thus been proposed to arrive at suitable simple analytic solutions. For example,   \citet{EbertAlbrand:1963:The-applicability_a} proposed to integrate eq.~(\ref{eq:diffthermaltrans}) with \( \Theta (d/\lambda) = \left( 1 + d / \lambda \right)^{-1} \)  by ignoring the pressure and temperature dependence of \( \Theta \), after having noted that Knudsens expression shows the right limiting behaviour in both pressure regimes. In a series of papers \cite{Weber:1932:Zur-Theorie_a,WeberKeesom:1932:Experimentelle_a,WeberSchmidt:1936:Experimentelle_a}, Weber and coworkers developed a semi-empirical expression for  \( \Theta (d/\lambda) \)  that would be valid all over the pressure range, capturing all but a weak pressure dependence that needed to be added as a small correctional term. Still, the solutions were too cumbersome for  practical applications \cite{Liang:1951:Some_a,Miller:1963:On-the-Calculation_a}.    

Another approach to the problem has thus been to search for a simple step function that would directly describe the transition between viscous and molecular flows in terms of the pressure and temperature ratios at the two sides of a cylinder subject to a temperature gradient. Whether based on purely empirical grounds \cite{Liang:1951:Some_a,Liang:1953:On-the-calculation_a,BennettTompkins:1957:Thermal_a} or based on an approximative solution to eq.~(\ref{eq:diffthermaltrans}), many of the proposed  expressions took the following form  \cite{Liang:1951:Some_a,Liang:1953:On-the-calculation_a,BennettTompkins:1957:Thermal_a,Miller:1963:On-the-Calculation_a}, 

  \begin{eqnarray}   
 	\label{eq:thermaltrans} R-1 &=& \theta (x)  \left( R_\mathrm{K} -1 \right)\nonumber \\
\mathrm{with\ }\theta (x)  &= &\left(\alpha x^2+\beta x+f(x)\right)^{-1}\,,
 \end{eqnarray}
linking the relative difference in pressure to the relative deviation of the square root of temperature  \cite{Miller:1963:On-the-Calculation_a}. In this equation, \( R \) and \(R_\mathrm{K}  \) are the pressure and temperature ratios as defined previously (eqs.~(\ref{eq:DefPRat}) and (\ref{eq:KnudsenRat})), \( x \)  is a variable proportional to pressure (\( p_2 \)) that may depend on temperature (\( T_1,\ T_2 \)) and $\theta$ is another step-like function, necessary related but not identical to $\Theta$ in eq.~(\ref{eq:diffthermaltrans}).   \( \alpha \) and \( \beta \) are semi-empirical parameters,  and \( f(x) \) is a slowly varying function in \( x \) with \(\lim_{x \rightarrow 0} f(x) = 1 \), thus assuring the correct low pressure limiting behaviour. The correct high pressure limit  is automatically warranted by the functional form of $\theta (x)$ as long as ($\alpha \neq 0 \vee \beta  \neq 0 $). We note in passing that the most simple equation of the above type with \( f(x)=1 \) is due to Liang \cite{Liang:1953:On-the-calculation_a} and that \(  \theta (x)\) has been termed degree of thermal transpiration \cite{TakaishiSensui:1963:Thermal}.

 Other formulations, such as the Kanki$-$Iuchi$-$Kosugi (KIK) \cite{KankiIuchi:1976:Flow_a} equation, which is dressed as a power law between \( R \) and  \( R_{\mathrm{K}} \) or the Weber \cite{WeberSchmidt:1936:Experimentelle_a} and the Kavtaradze \cite{Kavtaradze:1954:Vliyanie_a} equations take different forms. These will not be presented and discussed in detail, but are given in~\ref{sec:TTE}  for reasons of completeness. The reason why we primarily concentrate on approaches conforming to eq.~(\ref{eq:thermaltrans}) is that three of these equations have already been demonstrated to be more accurate (better than 0.5\,\%)  than others in describing thermal transpiration effects in CDGs over the pressure range between 1 and 130 Pa -- at least when  N\( _2 \) is measured under ambient conditions \cite{YoshidaKomatsu:2010:Compensation_a}. As we will see later, this also holds for our measurements on argon. A summary of the different equations that we present in detail below can be found in Table~\ref{tab:thermaltranspiration}.
 \subsection{Takaishi and Sensui (TS) equations}\label{sec:TakSens}        

Probably the most commonly used parameterisation for describing the pressure dependence of thermal transpiration has been introduced by \citet{TakaishiSensui:1963:Thermal}:
\begin{equation}
	\label{eq:tseq} \theta (x) = \left(\alpha x^2+\beta x+\gamma \sqrt{x}+ 1\right)^{-1}\,
\end{equation}
	with \( x = 2 p_2 d/(T_1+T_2)\), and three constants \( \alpha, \beta \) and \( \gamma \)  which depend on the gas (values for Ar given in Table~\ref{tab:thermaltranspiration}), on temperatures \( T_1,\ T_2 \) and the pressure \( p_2 \).\footnote{In their article \cite{TakaishiSensui:1963:Thermal}, Takaishi and Sensui used capital and arabic letters to denote parameters and the  pressure dependent variable \( x \), but we have opted to return to the notation introduced previously (see Refs.~\cite{Liang:1953:On-the-calculation_a, WeberSchmidt:1936:Experimentelle_a, Miller:1963:On-the-Calculation_a}).} Thus, \( f(x)= \gamma \sqrt{x}+1 \) in eq.~(\ref{eq:thermaltrans}).

The three gas dependent parameters \( \alpha, \beta \) and \( \gamma \) need to be determined experimentally. \citet{TakaishiSensui:1963:Thermal} tested their equation on measurements of He, Ne, Ar, Kr, Xe, H\( _2 \), N\( _2 \), and CH\( _4 \) and found the following dependence on the molecular diameter \( D \): 
\begin{eqnarray}
 \alpha &=& 0.79 \exp(0.0117 D/\mathrm{pm})\, \mathrm{(mm\,Pa/K)^{-2}} \label{eq:TSCoefA},\\
 \beta &=& 0.042 \exp(0.014 D/\mathrm{pm}) \,\mathrm{(mm\,Pa/K)^{-1}} \label{eq:TSCoefB}, \\
 \gamma &=& \left(953\, \mathrm{pm} / D-1.21   \right) \, \mathrm{(mm\,Pa/K)^{-1/2}}, \label{eq:TSCoefC}
\end{eqnarray}           
where \(D  \) is  obtained from  viscosity data
\begin{equation}   \label{eq:diam}
	  \eta = \frac{5}{16 D^2}\sqrt{\frac{m k T}{\pi}},
\end{equation}  
and where the symbols \( m \) and \( k \) take the usual meanings of    molecular mass and the Boltzmann constant. 

Because of lack of theoretical basis, \citet{TakaishiSensui:1963:Thermal} advised careful use of equations~(\ref{eq:TSCoefA}) -- (\ref{eq:TSCoefC}). In particular the diameter dependence of \( \gamma \) 
seems to be questionable. First, as already pointed out by the authors, \( \gamma \) does not depend linearly on \emph{D}, which would be expected if thermal transpiration scales with the Knudsen number. Secondly, we note that \( \gamma \) becomes negative at values above 790\,pm, which predicts  different pressure  dependencies for  large and small molecules. Still, reasonable  agreement had been found using SF\( _6 \) (\( D\simeq600\, \)pm) and in the absence of a set of coefficients for a particular gas, application of the above formulae has generally been recommended \cite{PoulterRodgers:1983:Thermal,JitschinRohl:1987:Quantitative,Jitschin:1990:Accuracy_a,HylandShaffer:1991:Recommended_a,Jousten:1998:Temperature_a}  

	\citet{Yasumoto:1980:Thermal_a} included many more and larger molecules in his study and inferred a different set of \( \alpha \) and \( \gamma \) coefficients for the TS equation~(\ref{eq:tseq}). The measurements implied a linear dependence of \( \gamma  \) on the molecular diameter \( D \). Unfortunately, no explicit formula for  \( \beta \) could be determined and only some range has been specified: 
 	\begin{eqnarray}
	 \alpha &=& 2.2 \cdot10^{-9} (D/\mathrm{pm})^4 \ \,\mathrm{(mm\,Pa/K)^{-2}} \label{eq:YCoefA} \\
	 \beta  &=& 0.75\dots6.0 \ \mathrm{(mm\,Pa/K)^{-1}} \label{eq:YCoefB},  \\
	 \gamma &=& \left( 0.024 D/\mathrm{pm}-4.8   \right) \ \mathrm{(mm\,Pa/K)^{-1/2}}  \label{eq:YCoefC}           .
	\end{eqnarray}    
 In addition, as already stated by the author of the same study, the derived diameter dependency does only partly reproduce the experimental values. The agreement is particularly  limited  for the rare gases and the largest molecules.  
\begin{table*}[ht] \caption{\label{tab:thermaltranspiration} Comparison and parameterisation of thermal transpiration curves (eq.~(\ref{eq:thermaltrans}))} \small
	\vskip4mm \centering 
	\begin{tabular}{lllllllll} \hline Equation& \( f(x) \)&Gas & \( \alpha \) & \( \beta \) &\( \gamma \)& \( \delta \)&\(x\textsuperscript{a} \)&Ref. \T\B \\
		\hline 
		TS (\ref{eq:tseq}) & \( \gamma \sqrt{x} +1\)&Ar\textsuperscript{b}&\( 60.8\)\,(mm\,Pa/K)\textsuperscript{-2}&\( 6.06\)\,(mm\,Pa/K)\textsuperscript{-1}& \( 1.35\)\,(mm\,Pa/K)\textsuperscript{-1/2}&---&\( p_2 d\left/\,\overline{T}\right. \) &\cite{TakaishiSensui:1963:Thermal} \T \\                      
		 Y-TS (\ref{eq:tseq}) &\( \gamma \sqrt{x} +1 \) &Ar\textsuperscript{b} &\( 50.6 \)\,(mm\,Pa/K)\textsuperscript{-2}&\( 5.25 \)\,(mm\,Pa/K)\textsuperscript{-1} &\( 4.33 \)\,(mm\,Pa/K)\textsuperscript{-1/2}&---&\( p_2 d\left/\,\overline{T}\right. \) &\cite{Yasumoto:1980:Thermal_a}\\                                      
		\v{S}etina (\ref{eq:tseq}) &\( \gamma \sqrt{x} +1\)&all\textsuperscript{c}&0.0293 &0.292&0.238&---&\( p_2 d\left/\, \overline{\eta v_{th}} \right.\)&\cite{Setina:1999:New-approach_a} \\  
		Miller (\ref{eq:miller}) &\( (1+\gamma x)/(1+\delta x) \)&all\textsuperscript{d}& 3/100 & 245/1000 & 5/2 & 2 &\( d/\lambda = p_2 d \pi \sqrt{2} D^2/(k \overline{T})\) &  \cite{Miller:1963:On-the-Calculation_a} \B \\  
		\hline \multicolumn{6}{l}{\textsuperscript{a} barred values are evaluated at the mean temperature \( \overline{T}=(T_1 + T_2)/2 \). } \T \\  
		\multicolumn{6}{l}{\textsuperscript{b} after conversion to SI units. Original values were based on pressure values in Torr. } \\
		\multicolumn{6}{l}{\textsuperscript{c} based on measurements of Ar, H\( _2 \), He, and N\( _2 \). } \\   
		\multicolumn{6}{l}{\textsuperscript{d} based on measurements of H\( _2 \), He, Ne, Ar,  Kr and  Xe.} \\
	\end{tabular}
\end{table*}

Based on a study on the four gases He, Ne, Ar, and N\( _2 \), \citet{Setina:1999:New-approach_a} found that the TS equation~(\ref{eq:tseq}) can be cast into universal form, \emph{ie} can be applied to all gases using a unique set of parameters (see Table~\ref{tab:thermaltranspiration}). This could be achieved through introducing a normalised pressure scale \(x = p_2/ p^\star \), where the characteristic pressure is given by
\begin{equation} \label{eq:pstar} 
	  p^\star  = \frac{\overline {\eta}\, \overline{v_{th}}}{d} =     \frac{5}{4\sqrt{2}}\frac{k \overline{T}}{d \pi D^2}  
\end{equation}
 and where \(  \overline{\eta} = \eta\left( \overline{T}\right)\) and \(  \overline{v_{th}} = \sqrt{8 k \overline{T} /(m \pi)}\) denote viscosity and  mean thermal molecular velocity at the average temperature $\overline{T}=(T_1+T_2)/2 $. 
 In this approach, the gas dependence is fully contained in the characteristic pressure \(  p^\star \), which via eq.~(\ref{eq:diam}), can be expressed in terms of the kinetic molecular diameter \( D \). 
 Note, however, that the definition of \v{S}etina \cite{Setina:1999:New-approach_a}, as well as the use of the formulae of Takaishi and Sensui \cite{TakaishiSensui:1963:Thermal}
  or Liang \cite{Liang:1951:Some_a} by Poulter et al.
   \cite{PoulterRodgers:1983:Thermal} 
   or by Jitschin and Rhl 
    \cite{JitschinRohl:1987:Quantitative} are not entirely consistent with the
   original definitions \cite{TakaishiSensui:1963:Thermal,Liang:1951:Some_a}, even though eq.~(\ref{eq:tseq}) takes an identical form. This is because these authors use $R = p_2/p_1$ instead of $p_1/p_2$, despite the fact that $T_1 < T_2 $ in their studies. With the coefficients listed in Table~\ref{tab:thermaltranspiration}, the half pressure \( p_{1\ \!\!\!/2} \), where the thermal correction reaches half of the Knudsen limit, \( (R-1)/(R_{\mathrm{K}}-1) =1/2 \), is about twice the value of  \( p^\star    \) 
 \begin{equation}   \label{eq:SetHalfP}
	   p_{1\ \!\!\!/2}  = 1.923\, p^\star        
\end{equation}                   
and can be calculated for each gas from viscosity data, either obtained experimentally or estimated from critical parameters \cite{GreenPerry:2008:Perrys_a,PolingPrausnitz:2001:The-Properties_a}.
                                        
\subsection{Miller equation}\label{sec:Miller}    
Already in 1963, \citet{Miller:1963:On-the-Calculation_a}   has proposed a universal equation as an approximate solution to the differential equation of \citet{WeberSchmidt:1936:Experimentelle_a} which contained the term \( f(x)=(1+ \gamma x)/(1+\delta x) \):
\begin{equation}
	\label{eq:miller} \theta (x) = \left(\alpha x^2+\beta x+\frac{1+\gamma x}{1+\delta x}\right)^{-1} \,, \qquad T_1 < T_2 
\end{equation}    
where the coefficients \( \alpha = 3/100, \beta= 245/1000, \gamma = 5/2 \) and \( \delta=2 \) have been determined as a ''best fit'' to experimentally available data on H\( _2 \) and the rare gases He, Ne, Ar,  Kr and  Xe and where the pressure dependent variable 
\begin{equation}     \label{eq:doverlambda}
x = d/\lambda = p_2 d \pi \sqrt{2} D^2/(k \overline{T}) 
  \end{equation}    
is the inverse Knudsen number. With eq.~(\ref{eq:diam}) and the mean thermal velocity \( \overline{v_{th}}  \) as defined above, 
we readily obtain 
\begin{equation}
	\label{eq:xmiller}
 x = \frac{5}{4} \frac{p_2}{p^\star} =  \frac{5}{4} \frac{\overline {\eta}\, \overline{v_{th}}}{d} .
\end{equation}
 This provides a normalised pressure scale, which seems to be shifted by 20\,\% as compared to the one introduced by \v{S}etina,
but the half pressure determined by eq.~(\ref{eq:miller}) 
\begin{equation}  \label{eq:MillerHalfP} 
	  p_{1\ \!\!\!/2}  = 1.983\, p^\star        
\end{equation}  
differs only by 3\,\% from the value for the half pressure of the \v{S}etina equation~(\ref{eq:SetHalfP}), indicating that both transition curves are indeed closely situated.
            From Table~\ref{tab:thermaltranspiration} it becomes clear that the linear \(  \beta x\) term is very similar to the one obtained by \v{S}etina, when the scaling factor of 5/4 is taken into account. At higher pressures, however, the Miller curve should fall off  somewhat more  rapidly than the \v{S}etina equation due to the values of \( \alpha \) being almost identical and the pressure scale being shifted by 20\,\%.  
			
			Most of the gas dependence is already contained in the variable $x$, but the coefficients $\alpha$ and $\beta$ also are slightly gas dependent. Their gas dependence can be inferred from an analysis of the underlying work of \citet{Weber:1932:Zur-Theorie_a}, where the gas flow has been derived from continuum mechanics using the following slip boundary condition \cite{Sharipov:2011:Data_a}
   \begin{equation}\label{eq:vel}
 u_y=\sigma_p 2 \sqrt{\frac{M}{R_gT}}\frac{\mu}{\rho}\frac{\mathrm{d} u}{\mathrm{d} x}+\sigma_T \frac{\mu}{\rho T}\frac{\mathrm{d} T}{\mathrm{d} x},
    \end{equation}
	where $u_y$ is the gradient of the flow velocity, $M$, $R_g$ and $T$ respectively are molar mass, the universal gas constant and the temperature, and where $\mu$ and $\rho$ are  viscosity and density. 
	$\mathrm{d} u/\mathrm{d} x$ and $\mathrm{d} T/\mathrm{d} x$ are the velocity and temperature gradients along the wall coordinate $x$. The viscous and temperature slip coefficients $\sigma_p$ and $\sigma_T$ are gas dependent proportionality factors, that can be inferred from experiments. The review of \citet{Sharipov:2011:Data_a} also provides a summary of experimental data. 
	The first term describes the hydrodynamic viscous flow driven by a pressure gradient. The second term describes the thermal creep due to a temperature gradient. 
	A comparison with the derivations by \citet{Miller:1963:On-the-Calculation_a} and \citet{Weber:1932:Zur-Theorie_a} yields 
	\begin{eqnarray} \label{eq:GasdepMillera}
		\alpha , \beta , \gamma  &\propto& \sigma_T^{-1},	 \\ \label{eq:GasdepMillerb}
	\beta , \gamma & \propto& \sigma_p .
	\end{eqnarray}

 \subsection{Gas dependence}\label{sec:GasDep}         

\begin{figure}[h!]
	\includegraphics[width=8.5cm]{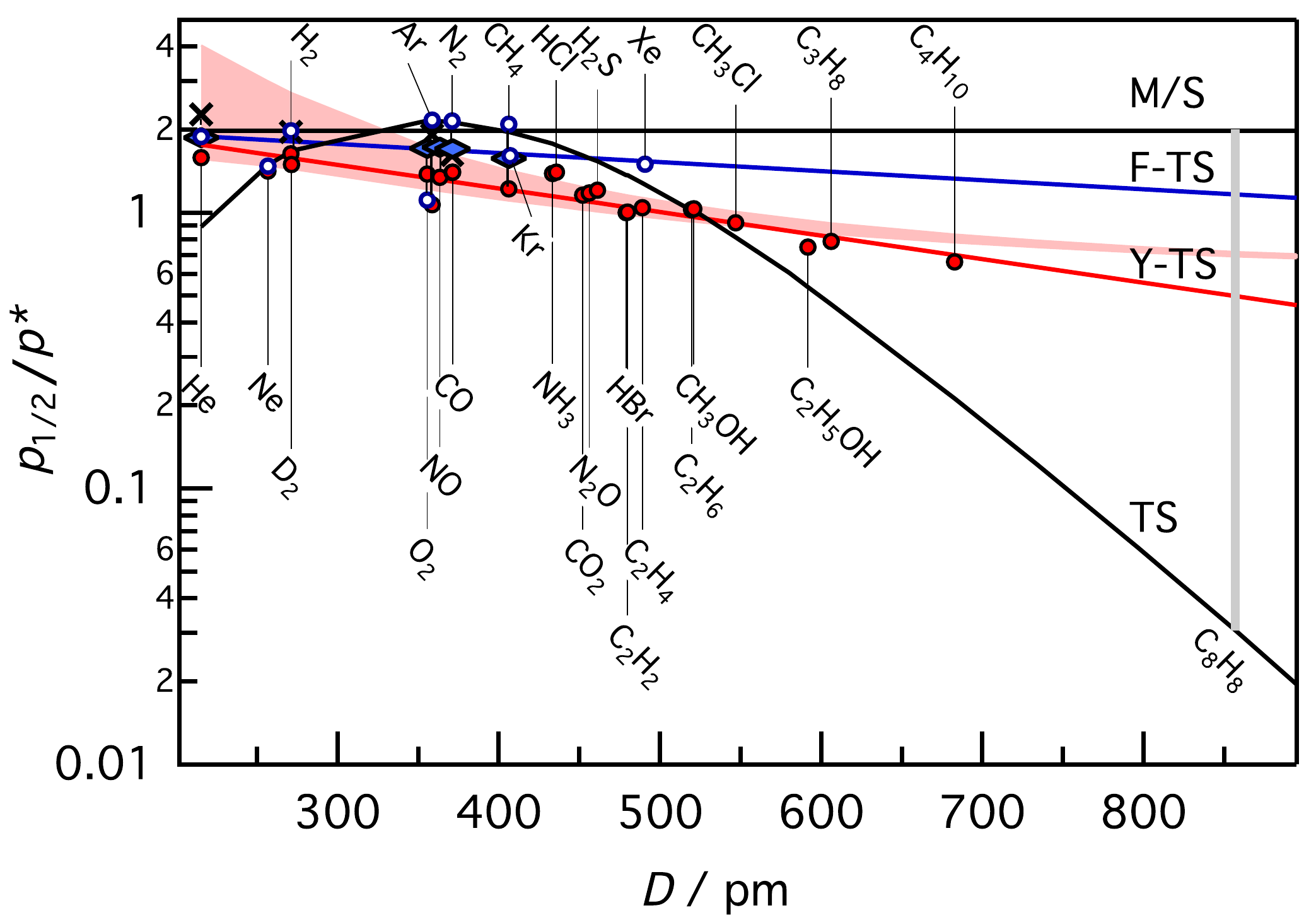} \caption{\label{fig:gasdep} Gas dependence of different thermal transpiration parameterisations indicated by the dependency of the half pressure (\( p_{1\ \!\!\!/2} \))  on the molecular diameter (\( D \)). Half pressures are given in units of the characteristic pressure \( p^\star \) (eq.~\ref{eq:pstar}). Black line M/S -- Miller equation and \v{S}etina equation; Black curve (TS) -- TS equation; Shaded (red) area (Y-TS) -- prediction by the Yasumoto modification of the TS parameterisation; symbols correspond to measurement data from Refs.~\cite{TakaishiSensui:1963:Thermal} (open circles), \cite{Yasumoto:1980:Thermal_a} (closed circles),  \cite{Setina:1999:New-approach_a} (diagonal crosses), and \cite{Furuyama:1977:Measurements_a} (rhombuses). 
	Straight lines are fits to the data of Yasumoto (Y-TS) and Furuyama (F-TS). Viscosities have been taken from Refs.~\cite{GreenPerry:2008:Perrys_a} and \cite{BichMillat:1990:The-viscosity_a}. The grey vertical line indicates the range of half pressures predicted for the styrene (C\( _8 \)H\( _8 \)) molecule.} 
\end{figure}
Table~\ref{tab:thermaltranspiration}  gives an overview of the specificities of the treated thermal transpiration equations. While the functional form of the curves is quite similar, the first two equations fundamentally differ from the latter by their gas dependence. As discussed before, \( \gamma \)  in the TS equation does not at all scale with  \( d/ \lambda \) and in the Yasumoto modification \( \gamma \) is a linear function in \( d/ \lambda \), but still has a non-zero offset. The normalised pressure where the degree of thermal transpiration (eq.~(\ref{eq:thermaltrans})) equals 1/2 therefore depends on the molecular diameter. Figure~\ref{fig:gasdep} shows the comparison of the four different dependencies. Neglecting the slight gas dependence inherent in the slip coefficients (eqs. (\ref{eq:GasdepMillera}) and (\ref{eq:GasdepMillerb})), the Miller and the \v{S}etina normalised half pressures \(p_{1\ \!\!\!/2}/ p^\star \simeq 2\) are independent of the molecular diameter and cannot be distinguished on the logarithmic scale in Fig.~\ref{fig:gasdep}. The TS equation (\ref{eq:tseq}), however, shows a very different dependence on the molecular diameter and roughly agrees with the previous two equations only for small molecules in the 250$-$500\,pm range.  With increasing diameter, normalised half pressures  \(p_{1\ \!\!\!/2}/ p^\star\) become smaller and for styrene (\( D \simeq  860\,\)pm) there is already a factor of 70 difference as compared to the Miller or the \v{S}etina predictions. Again, it should be pointed out here, that this mismatch is entirely due to extrapolation of eqs.~(\ref{eq:TSCoefA}) -- (\ref{eq:TSCoefC}), that have been obtained from a fit on data over a restricted range. As can be seen from Fig.~\ref{fig:gasdep}, the measurement data itself does not necessarily support the dependence inherent in these equations. The modified TS-Y equation, on the other hand, shows a comparatively weaker gas dependency. However, its transition pressures for small molecules ($D \lesssim 500$\,pm) -- and this holds particularly for the measurements -- are generally lower than the predictions of the other three parameterisations.

\section{Experimental and method}     \label{sec:exp}

\begin{figure}[floatfix]
	\begin{center}
	\includegraphics[width=5cm]{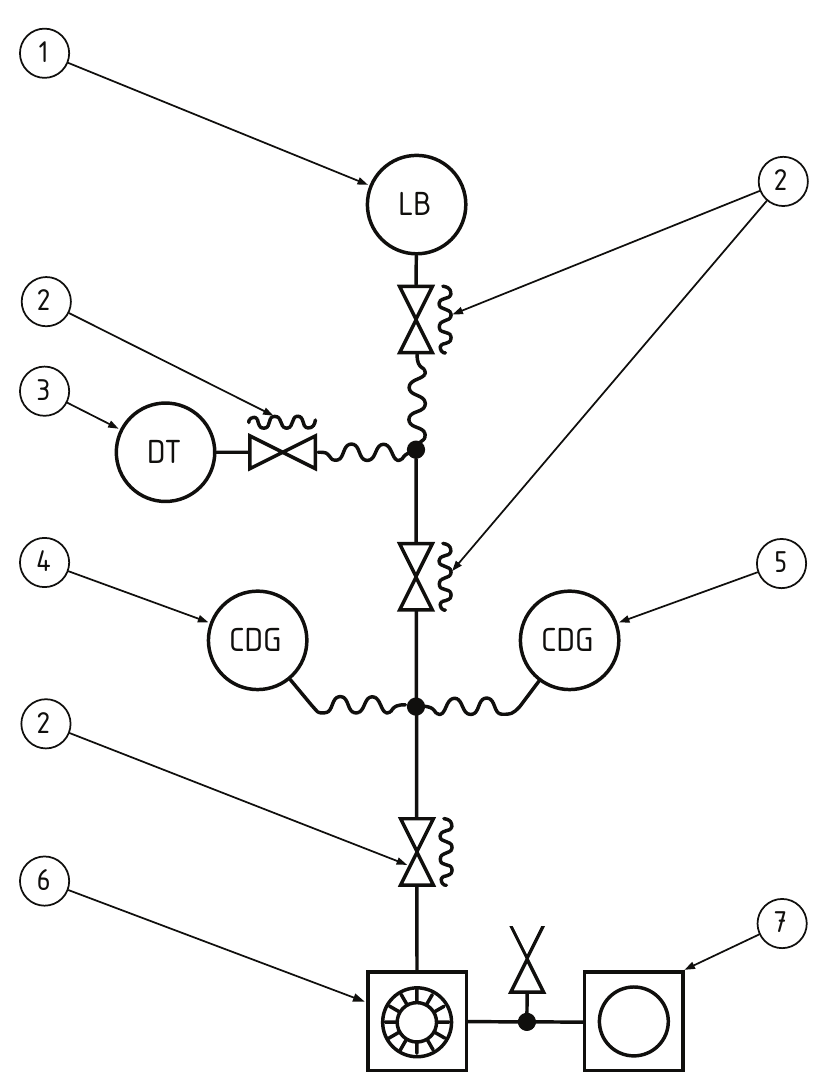} \caption{\label{fig:vacsys} Vacuum setup. Argon is stored in a lecture bottle (1) and can be added to the system via a stainless steel bellow sealed valve (2). Styrene, of which the vapour can be fed to the system, is kept as a liquid in a stainless steel dip tube (3). The gas inlet lines connect to two commercial capacitive diaphragm gauges with 133 Pa (4) and 1.33 kPa (5) full range via an additional bellow valve. Both CDGs are linked to adapted controller readouts. The central volume can be evacuated by a turbomolecular pump (6) backed up by a diffusion pump (7). All lines are made out of 6.25\,mm  diameter stainless steel tubing, except for the connection to the turbomolecular pump, which consists out of a 40\,mm inner diameter bellow. } 
	\end{center}  
\end{figure}

Fig.~\ref{fig:vacsys} depicts the experimental setup, consisting out of a gas feeding line, the pressure sensors and the turbomolecular pumping system. The two CDGs are a 1.33\,kPa head (CDG1, model 390, MKS Instr.) connected to a model 270 B-4 readout (MKS Instr.) and a 133\,Pa gauge (CDG2, model 690, MKS Instr.) linked to a type 670 controller (MKS Instr.). Before the experiments, proper operation of gauge heads and controls at 296 K as well as their compliance with manufacturer specifications has been verified by MKS France using a certified (DKD/DAkkS) instrument for comparison. To minimise the effect of ambient temperature variations, the pressure sensors have been protected by several layers of insulating bubble wrap. Gas supplies were laboratory grade argon (Alphagaz~1, 99.999\,\% purity) from Air Liquide (France), which was used without further purification, and styrene that was acquired from Sigma Aldrich (Germany) with a purity of better than 99\,\%. The liquid has been filled into a stainless steel dip tube under an argon atmosphere and has been subjected to several freeze and thaw cycles before the measurements. 

Following the work of \citet{Baldwin:1973:Thermal_a} and \citet{PoulterRodgers:1983:Thermal}, we chose to determine the thermal transpiration effect by comparing a heated CDG, operating at standard temperature \( T_2 = 318.15\,\)K, with an unheated one that was kept at \( T_1\simeq 293 \dots 300\, \)K. Void of a temperature gradient, the unheated head does not suffer from thermal transpiration, but due to being operated out of specifications its readings cannot be trusted right away and an in-situ calibration is required. 

Three configurations have been necessary to establish the measurement procedure, which aimed at minimising measurement uncertainties by performing relative rather than absolute pressure measurements. Firstly, systematic and possibly pressure dependent biases between the two sensors have been determined in simultaneous pressure measurements operating the sensors as described below. Despite the presence of temperature gradients within the two gauges, differences in the observed signals are largely due to controller or gauge specific characteristics, such as capacitor non-linearity \cite{JoustenNaef:2011:On-the-stability_a} or controller gain and offsets. Secondly, it was verified that these sensor specific characteristics do not depend on whether the sensor is heated or not. For that purpose, the two gauges have been operated without heating and the relative deviation between the two sensor readings has been calculated after correcting for the relative gauge sensibilities determined in the first step. Finally, thermal transpiration measurements on argon and styrene have been performed with one gauge heated and the other not.    

\begin{figure}                           
	\begin{center}
	\includegraphics[width=8.5cm]{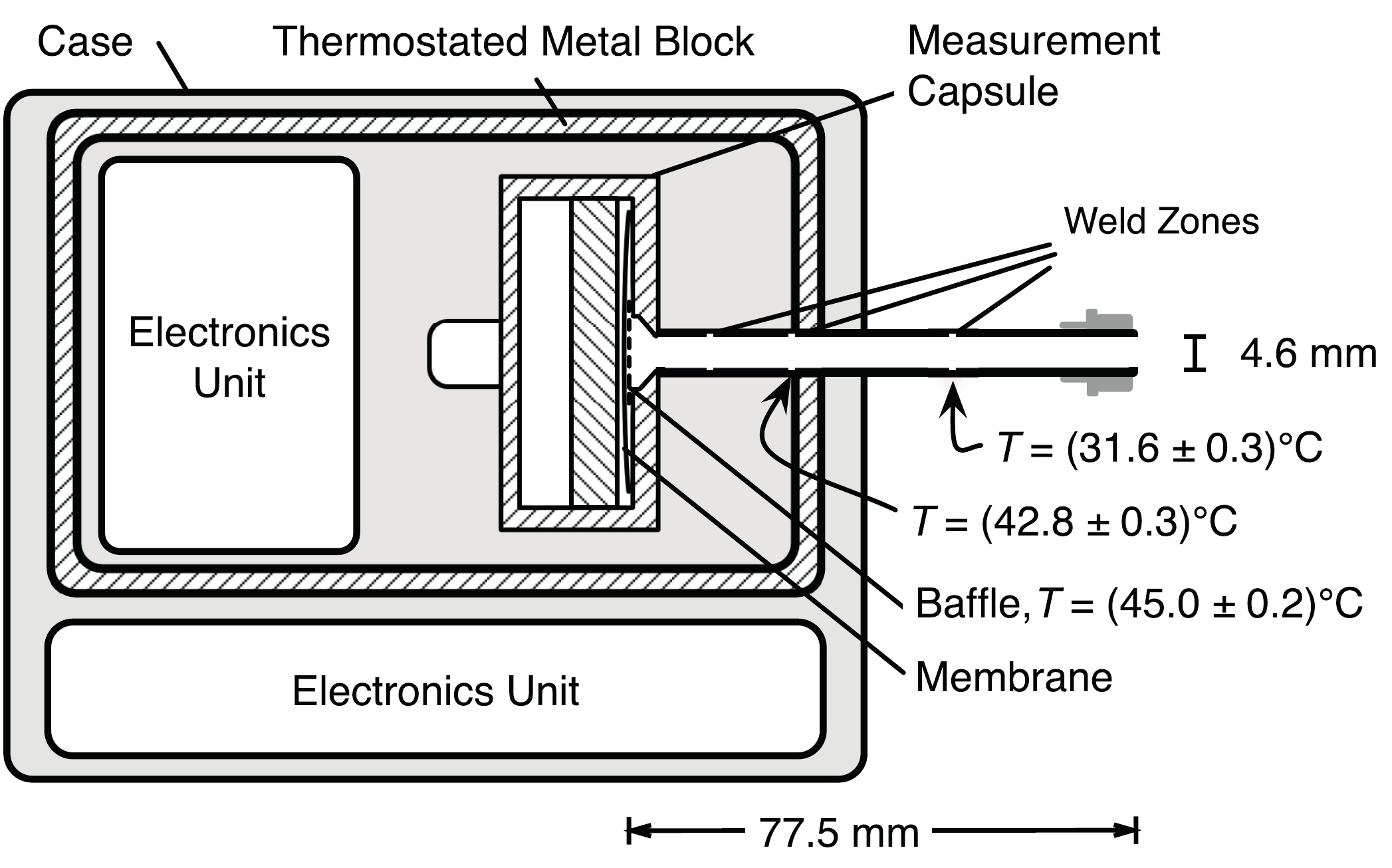} 
	\caption{\label{fig:BaratronGeometry} Principal scheme of the gauge head which is approximately to scale, except for the diaphragm where dimensions have been exaggerated. Dimensions were taken from the literature and/or confirmed by own measurements on an opened 390 sensor. The VCR\textsuperscript{\textregistered} fitting (on the right) connects to the sensor via a 4.6\,mm inner diameter stainless steel tube. The diaphragm is mounted into a cylindrical capsule, which is placed inside a thermostated metal block. The baffle temperature of $(45.0 \pm 0.2)\,^\circ $C has been confirmed by measurement. Temperatures were determined using both a calibrated thermocouple and a PRT. The values give expanded standard uncertainties with \( k=2 \). Note that the connecting tube has not a simple cylindrical geometry. About 4 mm from the baffle, which protects the membrane, the tube becomes conical and opens up to a thin, 11 mm diameter wide cylindrical volume. The 4.6 mm diameter tube itself seems to be made out of four distinct pieces connected by welds which create zones of increased inner diameters. }  
 \end{center}
\end{figure}

Sensor temperatures have been determined after the transpiration measurements have been terminated. The heated CDG temperature  \( T_2 \) was measured using a calibrated thermocouple and a calibrated platinum resistance thermometer (PRT-100), which were inserted through the vacuum connector tube after venting the instrument. A continuous gradient has been observed along the 4.6\,mm inner diameter tubing whose dimensions were provided by the manufacturer and have been verified by calliper measurements. The principle geometry and characteristic temperatures are shown in Fig.~\ref{fig:BaratronGeometry}. While the nominal temperature of about \( T =318\,{} \)K have been confirmed at the inner part of the sensor, a roughly 2\,K lower temperature \( T = (315.95 \pm 0.3)\,\)K has been measured just at the inner edge of the thermostated metal block. We found that temperatures of the second gauge were within \( 0.1 \,{}\)K of those of the first one, when it was thermostated. This indicates that our observed temperature distributions are somewhat representative. 

The low temperature \( T_1 \) in the unheated CDG has been determined from  temperature readings at the outside of the CDG. These values have been corrected by an empirical offset \( \Delta T_1 = (0.6 \pm 0.1)\,\)K due to the heating caused by the gauge electronics with the heater switched off. As for the heated CDG, the offset has been determined from thermocouple and PRT-100 measurements at the open sensor. We further note that the notion of a plain cylindrical tubing connecting the heated sensor compartment and the VCR connector is too simplistic. Inspection of the opened 390 HA gauge shows that three weld zones exist, where different tubes of equal inner diameter are connected. This leads to the creation of concentric gaps with increased inner diameter, the first about 28\,mm behind the gas entry and the two others further in the gauge. At about 4 mm from the baffle protecting the membrane, the tube opens up to an 11 mm wide cylindrical disk. While the apparent deviation from the cylindrical geometry alone might justify the use of an effective rather than the geometric diameter of the connector tube \cite{JitschinRohl:1987:Quantitative}, the small temperature differences within the heated metal block implying that the more complicated geometry close to the membrane does not largely contribute to thermal transpiration  within the CDG.

\section{Results and discussion} 

\subsection{Sensor calibration and uncertainties} 
The measurand $R=p_1/p_2$ (or, throughout this section, equally convenient its inverse $p_2/p_1$) is obtained from the pressure signals of the two CDGs after suitable correction for biases. These needed to be determined in calibration measurements, which also allowed to determine the measurement uncertainty.

\begin{figure}[floatfix]
	\includegraphics[width=8.5cm]{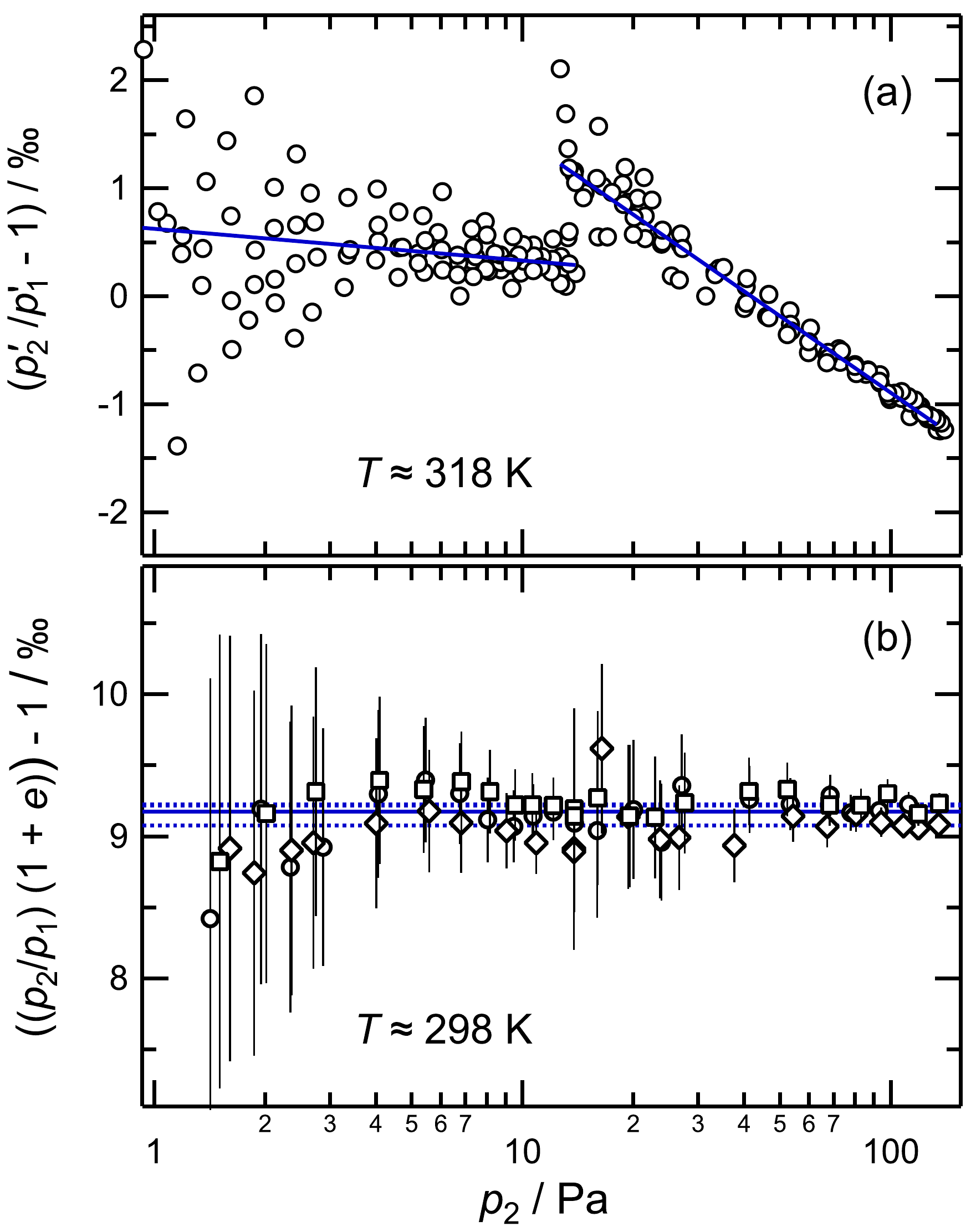} \caption{\label{fig:pcal} Relative difference in reading between two CDG sensors 1 (133 Pa FS) and 2 (1.33 kPa FS) with (a) and without (b) heating through the integrated sensor thermostat. 
(a) Best fits on the direct CDG readouts ($p^\prime_2/p^\prime_1-1$) have been established in two separate subranges 13.3 \( - \) 133\,Pa and \( < 13.3\, \) Pa, corresponding to different controller settings. (b) Relative deviation of two sensor pressures with unheated gauges at an ambient temperature of about 25\,\({}^\circ \)C after correcting for the systematic bias in (a). $p_1/p_2$ is the pressure ratio after correction of the pressure independent offset $e$ (see eq.~(\ref{eq:bias}) for a definition of all correction terms). Three series of measurements indicated by different symbols have been performed. Dotted horizontal lines are extrapolations from 3 to 4 highest pressure points (\( p> 70 \)\,Pa) of each series; the solid line averages on these three values. Uncertainties are given at a 95\,\% level of confidence.} 
\end{figure}

Figure~\ref{fig:pcal} shows the results of the calibration measurements that were done with argon as a working gas. The calibration under heated conditions in Fig.~\ref{fig:pcal}a yield the pressure dependent relative sensitivity of CDG2 vs CDG1, given by the relative deviation ($p_2^\prime/p^\prime_1-1$) of the two uncorrected signals from the sensor readouts.  Depending on the measurement range set by the controllers (133 Pa or 13.3 Pa), distinct sensitivities have been found. In the high pressure range (\(13.3  - 133\)\,Pa), the relative sensitivity changes gradually from \( -1.2 \) to \( +1.2 \)\,‰; in the low pressure range, there is a positive offset with a small possible trend. At our calibration temperature of 300\,K, relative deviations between the two sensors of \(\pm  1.4 \)\,‰  and \(\pm  3.5\)\,‰ at 133\,Pa and  13.3 Pa, respectively, are within the manufacturer specified range and indicate normal operation.

 Relative differences in the pressure readings $p_2^\prime/p^\prime_1-1$ reflect type A measurement uncertainties as well as systematic bias \cite{JCGM/WG-1-2008:2008:Evaluation_a}. While systematic bias is apparent from the trend lines given in Figure~\ref{fig:pcal}a, type A uncertainties, such as display resolution, reading uncertainties as well as offset and short-term instabilities, are indicated by the scatter and could be determined from analysing the residuals. The bias corrected pressure ratio can be fitted by
\begin{equation}\label{eq:bias}
		  \frac{p_2}{p_1}= \frac{p_2^\prime}{p_1^\prime}\bigl(1+ c_1 + c_2 \cdot \log \left(p/\mathrm{Pa}\right)\bigr)^{-1}\bigl(1+e\left(T_1, T_2\right)\bigr)^{-1},
\end{equation}
with $p_1^\prime$ and $p_2^\prime$ being the indicated pressure values on the two sensors CDG 1 and CDG 2, $c_1$ and $c_2$ specifying the pressure dependent corrections and $e$ a temperature dependent term, to be determined in an additional measurement. By definition $e=0$, if $T_1 = T_2 = 318.15\,$K. Because correction parameters, $c_1, c_2 $ and $e$ are small ($< 10^{-2}$), an eventual pressure dependence of $e$ and a temperature dependence of $c_1$ and $c_2$  can be neglected, which is also confirmed by the measurements displayed in Fig.~\ref{fig:pcal}b. From Fig.~\ref{fig:pcal}a, we determined
   $ c_1 = 0.00062385$ and $c_2 =-0.00029358$ for the low ($\leq 13.3$\,Pa) as well as   $ c_1 =  0.0038223$ and $c_2 = - 0.0023579$ for the high ($> 13.3$\,Pa) pressure range. 
   
    The scatter inherent in Fig.~\ref{fig:pcal}a reflects the uncertainties of individual measurements after bias correction. From the homoscedastic standardised residuals of the fits, relative standard ($k=1$) uncertainties of $u_r(p_2/p_1)=4.9\cdot 10^{-3} (p_2/\mathrm{Pa})^{-1}$ and $u_r(p_2/p_1)=1.2\cdot 10^{-3} (p_2/\mathrm{Pa})^{-1}$ have respectively been determined for the high and low pressure ranges.   
   
   The correction $e(T_1, T_2)$ due to change of sensor temperatures (see eq.~\ref{eq:bias}) must be determined for each temperature configuration ($T_1, T_2$). It can be inferred from high pressure measurements ($\sim 133$\,Pa), where thermal transpiration effects can be neglected. Using Ar and the \v{S}etina model as an example, we find the degree of thermal transpiration  $\theta( p>120\,\mathrm{Pa}) < 0.9\,$\%. Even for $T_1=298\,$K and $T_2=318\,$K, the associated bias on $p_2/p_1$ will thus be at the 0.3\,‰ level or below. It will be completely negligible for the measurements with smaller temperature differences and for the styrene measurements. 
   
    The applicability of the bias correction scheme in eq.~(\ref{eq:bias}) and the validity of the derived uncertainties have been confirmed through a second calibration keeping both sensors at ambient temperature. Fig.~\ref{fig:pcal}b shows the results with the data  already  corrected for the sensor sensitivities from Fig.~\ref{fig:pcal}a. The data are compared to constant offset values determined from the average of four to three highest pressure values, which vary between 9.08 and 9.22\,‰ for three different measurement series. 
	 Evidently, the data are compatible with a constant correction term, even though the curvature apparent in the low pressure data points towards a small, albeit non-significant residual bias. Assuming negligible relative pressure differences at high pressure ($\gtrsim 120\,$Pa), we can therefore measure relative pressures and pressure differences within the stated uncertainty of a few per mil over the pressure range from 1 to 133\,Pa as long as sensor temperatures are within $\sim 298$ and 318\,K. 
	 The corresponding standard ($k=1$) uncertainties are thus a combination of the previously determined individual Type A uncertainties and the uncertainty of the offset $e$, which has been determined to be $u(e)=3.0\cdot 10^{-5} $:
	\begin{eqnarray}\label{eq:uncertainties}
		u_r\left(p_1/p_2\right)&=&\sqrt{\left(4.9\cdot 10^{-3} \left(p_2/\mathrm{Pa}\right)^{-1} \right)^2 + \left(3.0\cdot 10^{-5}\right)^2} \nonumber\\
		&& \mathrm{for}\,\ \ 13.3\,\mathrm{Pa} < p_2 \leq 133\,\mathrm{Pa},\\
		%
		u_r\left(p_1/p_2\right)&=&\sqrt{\left(1.2\cdot 10^{-3} \left(p_2/\mathrm{Pa}\right)^{-1} \right)^2 + \left(3.0\cdot 10^{-5}\right)^2} \nonumber\\ 
&		&	\mathrm{for}\, \ \ p_2 \leq 13.3\,\mathrm{Pa}.
	\end{eqnarray}

\subsection{Thermal transpiration correction for argon}  \label{sec:argon}

Based on the calibration technique just outlined above, the thermal transpiration curve of argon has been determined in the pressure range between 1 and 130\,Pa. The temperature of the unheated Baratron was \( T_1 = (298.15 \pm 0.6) \)\,K (\( 25.8\,^\circ \)C), where the expanded  (\( k=2 \))  standard uncertainty essentially reflects changes of room temperature between different measurement series. Following \citet{JitschinRohl:1987:Quantitative}, we choose an effective temperature for the heated sensor of \( T_2 = 315.95 \,K (42.8\,^\circ \)C), which is smaller than the nominal temperature of $45.0\,^\circ$C. This choice provides the best match to our data, but implies a low pressure limit $R_0 > R_\mathrm{K}$. Because our measurements are restricted to ranges where this limit has not yet been reached, we cannot safely conclude that this is a significant result. Nevertheless, it needs to be pointed out that such an effect has been observed with helium on pyrex \cite{HobsonEdmonds:1963:Thermal_a,Hobson:1969:Surface_a}, which is attributed to non-diffusive wall scattering \cite{Siu:1973:Equations_a} that might also take place on polished stainless steel.

\begin{figure}[b!]
	\begin{center}
		\includegraphics[width=8.5cm]{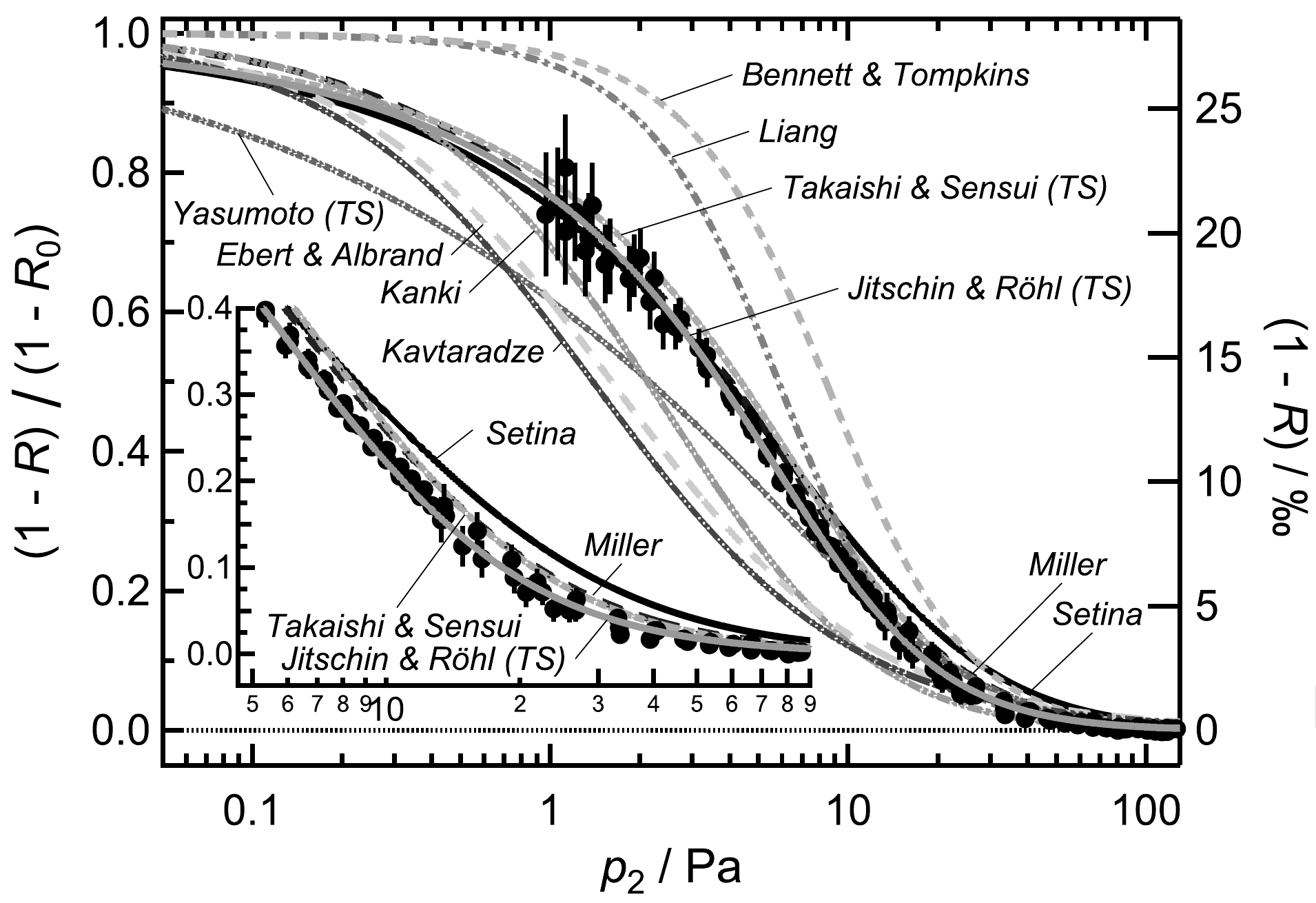} \caption{\label{fig:ThermTransArgon}  Thermal transpiration curve of argon as a function of pressure. Measurements (black circles) are plotted using the scale on the right axis ($1-R$). The effective degree of thermal transpiration  \(  \left(R-1\right)\left/\left(R_{\mathrm{0}}-1\right)\right. \) on the left axis has been obtained from setting \( T_1 = 298.95 \)\,K (\( 25.8\,^\circ \)C) and \( T_2 = 315.95\)\,K (\(42.8\,^\circ \)C). Measurements are compared to different models (lines) and bars on measurement data indicate expanded standard uncertainties for $k=2$. Model curves have been calculated using \( d=4.6 \)\,mm, except for the Jitschin and Röhl parameterisation of the TS curve where \( d=5.4 \)\,mm has been obtained as a fit result (see text). The inset provides a closer look at the onset of thermal transpiration for the four models that give best agreement with the measurements.}  
	\end{center}
\end{figure}

The results of the measurements are shown in Fig.~\ref{fig:ThermTransArgon}, along with available models and a more quantitative justification of our approach will be given at the end of this section.
 
As has been observed previously \cite{YoshidaKomatsu:2010:Compensation_a,JitschinRohl:1987:Quantitative}, the original and modified Liang equations show a too steep pressure dependence and a transition which is shifted towards higher pressures when compared to the measurements. The Kavtaradze, the Ebert$-$Albrand as well as the Kanki$-$Iuchi$-$Kosugi (KIK) equations are shifted towards lower pressures corresponding to half pressures \( p_{1\ \!\!\!/2} \) being smaller than the observation by factors between two and three.

The Yasumoto parameterisation for the TS model (referred to as Yasumoto (TS) in Fig.~\ref{fig:ThermTransArgon}) is also shifted towards low pressure and shows a much too weak pressure dependence -- likely because it is essentially derived from fits to data that essentially only cover the high pressure region (\( p > p_{1\ \!\!\!/2} \)). Indeed,  agreement with our measurements is reasonable at high pressures (\( \geq 10\)\,Pa). The predicted low pressure dependence, however, has been obtained from extrapolation and the previously demonstrated mismatch of their model curves with other data \cite{Yasumoto:1980:Thermal_a,Furuyama:1977:Measurements_a} is once more confirmed by our results. The important difference with respect to the TS model is due to the choice of parameters: Whereas \( \alpha \) and \( \beta \) in eq.~(\ref{eq:tseq}) are similar to those given by Takaishi and Sensui, \( \gamma \) is much higher. As a consequence, the  pressure dependence is weaker and the half pressure is smaller than that of Takaishi and Sensui. This is a general feature of the Yasumoto data, which show systematically low \( p_{1\ \!\!\!/2} \) values (see Fig.~\ref{fig:gasdep}). 

The best agreement with the measurements  is achieved either by the Miller formula (eq.~(\ref{eq:miller})), by the original
TS equation~(\ref{eq:tseq}) or by the modification proposed by \v{S}etina (see Tab.~\ref{tab:thermaltranspiration}), the latter being a little bit more off at the onset of the effect. This finding confirms earlier results on the thermal transpiration corrections for N\( _2 \) in CDGs \cite{YoshidaKomatsu:2010:Compensation_a}. Unlike this study on N$_2$, however, we find that all modelling curves, which are obtained without any parameter adjustment, are slightly offset towards higher pressure when compared to our measurements. This seems to be inline with other studies \cite{JitschinRohl:1987:Quantitative}, who already observed a slight transducer dependent discrepancy, using the equation and parameters proposed by Takaishi and Sensui to interpret  measurements on different gases.  As a solution, it was proposed to freely adjust the tube diameter \( d \) as well as the sensor temperature \( T_2 \).  If we thus adjust the  diameter as a free parameter, we obtain best agreement with our data with an effective diameter of \( d = 5.4\)\,mm, when \( T_2 \) is fixed to the value of $315.95\,$K. Consequently allowing for fitting of $d$ also in the Miller and Setina models yields best fit values of 5.2 and 5.3\,mm, respectively.

\begin{figure}
	\begin{center}
		\includegraphics[width=8.7cm]{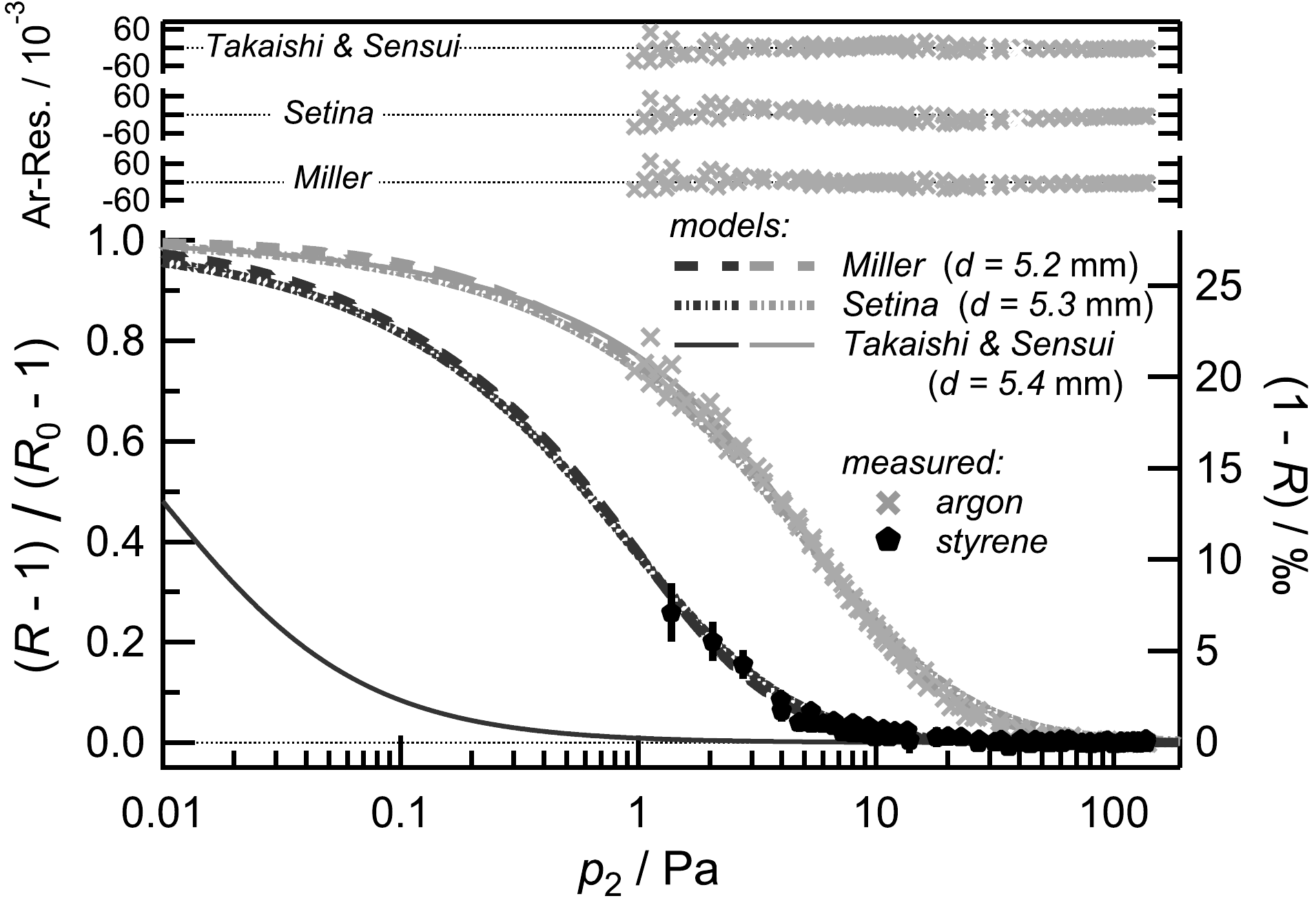} \caption{\label{fig:ThermTransArgonStyrene} Effective degree of thermal transpiration \(  \left(R-1\right)\left/\left(R_{\mathrm{0}}-1\right)\right. \) of argon (grey colour) and styrene (black colour) as a function of pressure using the effective temperature $T_2=315.95\,$K. 
		Measurements are compared to models of Miller, Takaishi \& Sensui and \v{S}etina using effective diameters of 5.2, 5.4 and 5.3\,mm, respectively. Vertical bars on styrene data indicate uncertainties on  the 95\,\% level of confidence. The styrene model curve of Takaishi \& Sensui has been calculated using equations (\ref{eq:TSCoefA}) -- (\ref{eq:diam}). Residuals for fits to argon data are given in the top traces.  }  
	\end{center}
\end{figure}

\subsection{Styrene and the gas dependency of thermal transpiration}   \label{sec:styrene}      
{\sloppy After having verified the calibration method and after having investigated the gauge characteristics using argon as a reference, the degree of thermal transpiration of styrene has been measured in the 1$-$130\,Pa range. The temperature of the styrene bath had been either (\( 298.15 \pm 0.3  \))\,K  or  (\( 299.35 \pm 0.3  \))\,K  during the measurements. }The result is displayed in Fig.~\ref{fig:ThermTransArgonStyrene} along with the measurements on argon. The figure also displays the different model curves using the effective diameters determined previously. It is evident that thermal transpiration in styrene occurs at pressures roughly 5 times lower than that of argon. The two models (\v{S}etina, Miller) which are based on a simple scaling of the ratio of tube diameter  over mean free path \( d/\lambda \propto \eta\, v_{th} \) (see eqs.~(\ref{eq:doverlambda}) and (\ref{eq:xmiller})) do satisfactorily describe the observed transpiration onset of styrene. On the contrary, the TS curve  predicts a much too small transition pressure of only \( p_{1\ \!\!\!/2} = 9\)\,mPa, which completely fails to match our data being more compatible with \( p_{1\ \!\!\!/2} = 0.63\)\,Pa, corresponding to $p_{1/2} = 1.8 p^\star$.      
Our data thus clearly demonstrate that the proposed extrapolation of parameters for large diameter molecules in eqs.~(\ref{eq:TSCoefA}) -- 
(\ref{eq:TSCoefC}) fails. It thus confirms the original suspicion of Takaishi and Sensui that the lack of a simple $Kn$ dependence in these equations is not correct. We therefore discourage from further using these equations. If thermal transpiration effects have to be estimated and predicted for yet un-investigated gases, use of either the Miller or the \v{S}etina equation is to be preferred. What is more, contrary to the formulae given by TS, the scaling of these two equations is  consistent with theoretical treatments. That such a scaling is also required from an experimental point of view is demonstrated here for the first time. 

Some uncertainty remains with respect to the role of gas specific interaction with the wall, where structural and material effects come into play.  It must be kept in mind that neither the viscous nor the thermal slip coefficient of styrene (and of all molecules that have not been studied so far) is known in advance and that theoretical modelling of the phenomena is particularly difficult for large polyatomic molecules. This has an impact not only on the prediction of $p_{1/2}$, but might also affect the low pressure limit of $R$, which might be increased such that  $R_0 > R_K$. Investigation of both these effects will be challenging and is clearly beyond the scope of this work. It must be recognised that these low pressure investigations will require narrow capillaries and very accurate pressure sensors, because with increasing molecular sizes \( d/\lambda \propto \eta\, v_{th} \) is generally decreasing, shifting the transition pressures towards lower values. 

 Using the observed span of accurate molecular data of the temperature slip for glass surfaces between 0.89 and 0.99 \cite{PorodnovKulev:1978:Thermal_a,Sharipov:2011:Data_a} as a guide, it seems reasonable to assume that transpiration effects in CDGs can be predicted using the Miller equation with effective diameters between 0.9 and 1.1\,$d$, in order to obtain upper and lower limits on the transpirational pressure ratios. Nevertheless, given the sparsity of data on larger polyatomic molecules we advise caution and point out the need for additional measurements. Since our results shed some doubt on previously applied thermal transpiration corrections that were based on eqs.~(\ref{eq:TSCoefA})--(\ref{eq:TSCoefC}), we re-examine some of these data, which also illustrates conditions and applications where these corrections need to be applied.

\subsection{Implications}   \label{sec:Implications}      

Table~\ref{tab:earlierresults}, which is by no means exhaustive, presents experimental conditions and thermal transpiration pressure ratios for studies where these effects are likely important. We have selected examples that comprise vapour pressure, absorption cross section and scattering cross section measurements, where molecular dimensions were sufficiently large that the failure of eqs.~(\ref{eq:TSCoefA}) - (\ref{eq:TSCoefC}) becomes apparent. We also discuss ozone, because its fragility implies that thermal transpiration of the molecule cannot be studied directly.

\begin{table*}[ht] 
	\caption{\label{tab:earlierresults} Comparison of different thermal transpiration ratios $p_1/p_2$ in vapour pressure and cross section measurements}
	\vskip4mm \centering 
	\footnotesize
	\begin{tabular}{lcccd{0}d{0}d{1}d{2}d{3}ccd{1}c} 
		\hline 
		\rule{0pt}{2.6ex} \multirow{2}{*}{Gas}&Measurement & \multicolumn{1}{c}{Mol. Diameter $D$\ \textsuperscript{b}} & \( d \) & \multicolumn{1}{c}{$T_1$} & \multicolumn{1}{c}{$ T_2$} & \multicolumn{1}{c}{$p$} & \multicolumn{5}{c}{Relative Pressure Difference $1-R$ \textsuperscript{c}} 
		& \multirow{2}{*}{Ref.} \T \\\cline{8-12}\rule{0pt}{2.6ex}
		
& Type\textsuperscript{a}& \multicolumn{1}{c}{(pm)} &(mm) &\multicolumn{1}{c}{ (K)} &\multicolumn{1}{c}{ (K)} & \multicolumn{1}{c}{(Pa)} & \multicolumn{1}{c}{Ref.\textsuperscript{d} (\%)} &\multicolumn{1}{c}{TS (\%)} & Miller (\%)& Miller (Pa) & \multicolumn{1}{c}{\v{S}etina (\%) }& 
\rule[-1.2ex]{0pt}{0pt} \B \\ \hline
	\multicolumn{1}{l}{ozone\textsuperscript{e}}      & VP  &  462     & 20 / 4 &  87 & 318 & 5.3 & 1.25                                & 1.56 & 1.9  \dots 2.6    & 0.10 \dots  0.14  & 2.7  & \multicolumn{1}{r}{\cite{HansonMauersberger:1985:precision}}\T\\
	\multicolumn{1}{l}{ethylene}               		  & SCS &  489     & 11     & 337 & 318 & 0.2 & \multicolumn{1}{c}{$\ \sim\,-3.0$}  & -2.2 & $-2.3$\dots$-2.2$ &    -0.0045       & -2.3  & \multicolumn{1}{r}{\cite{BettegaSanchez:2012:Positron_a}}\\
	\multicolumn{1}{l}{ethane}                        & ACS &  520     & 4.6    & 197 & 318 & 9.2 & \multicolumn{1}{c}{--}              & 1.09 & 1.9  \dots 2.7    & 0.18 \dots  0.25  & 2.8  & \multicolumn{1}{r}{\cite{HarrisonAllen:2010:Infrared_a}} \\
	\multicolumn{1}{l}{ethane }                       & SCS &  520     & 11     & 338 & 318 & 0.2 &                          -3.1       &-2.1  & $-2.4$ \dots$-2.3$& $-0.005$          &-2.3  & \multicolumn{1}{r}{\cite{ChiariZecca:2013:Cross_a}}\\
	\multicolumn{1}{l}{1,4 dioxane}                   & SCS &  580     & 11     & 337 & 318 & 0.2 &                          -3.0       &-1.6  & $-2.2$ \dots$-2.1$& $-0.004$          &-2.1  & \multicolumn{1}{r}{\cite{ZeccaTrainotti:2012:Positron_a}}\\
	\multicolumn{1}{l}{acetone}                       & ACS &  685     & 4.6    & 195 & 318 & 7.1 & \multicolumn{1}{c}{--}              & 0.16 & 1.2  \dots 1.7    & 0.09 \dots 0.12   & 1.8  & \multicolumn{1}{r}{\cite{HarrisonAllen:2011:Infrared_a}} \\
	\multicolumn{1}{l}{benzene}                       & SCS &  734     & 11     & 297 & 373 & 0.2 & \multicolumn{1}{c}{$<10$}           & 1.7  & 8.0  \dots 8.5    & 0.016\dots 0.017  & 8.1  & \multicolumn{1}{r}{\cite{ZeccaMoser:2007:Total_a}}\\
	\multicolumn{1}{l}{cyclohexane}                   & SCS &  772     & 11     & 297 & 373 & 0.2 & \multicolumn{1}{c}{$<10$}           & 1.1  & 7.1  \dots 8.2    & 0.015\dots 0.016  & 7.8  & \multicolumn{1}{r}{\cite{ZeccaMoser:2007:Total_a}}\\
	\multicolumn{1}{l}{benzoic acid  }                & VP  &  893     & 17     & 310 & 423 & 0.45& \multicolumn{1}{c}{--}              & 0.14 & 5.0  \dots 6.0    & 0.022 \dots 0.027  & 5.6  & \multicolumn{1}{r}{\cite{MonteSantos:2006:New-Static_a}}\\
	\multicolumn{1}{l}{naphthalene  }                 & VP  &  939     & 4.6    & 268 & 318 & 0.4 & \multicolumn{1}{c}{--}              & 0.06 & 4.9  \dots 5.3    & 0.020 \dots 0.021 & 5.0  & \multicolumn{1}{r}{\cite{RuzickaFulem:2005:Recommended_a}}\\
	\multicolumn{1}{l}{naphthalene\textsuperscript{f}}&(VP) &  939     & 4.6    & 283 & 473 & 2.7 & \multicolumn{1}{c}{--}              & 0.09 & 5.0  \dots 6.4    & 0.13 \dots  0.17  & 6.2  & \multicolumn{1}{r}{\cite{RuzickaFulem:2005:Recommended_a}}\\
	\multicolumn{1}{l}{naphthalene\textsuperscript{f}}&(VP) &  939     & 4.6    & 283 & 323 & 2.7 & \multicolumn{1}{c}{--}              & 0.008 & 0.7  \dots 1.0    & 0.020 \dots  0.027  & 1.0  & \multicolumn{1}{r}{\cite{RuzickaFulem:2005:Recommended_a}}\\
	\multicolumn{1}{l}{naphthalene  }                 & VP  &  939     & 17     & 268 & 423 & 0.4 & \multicolumn{1}{c}{--}              & 0.10 & 7.1  \dots 8.7    & 0.028\dots 0.035 & 8.1  & \multicolumn{1}{r}{\cite{MonteSantos:2006:New-Static_a}}\\
	\multicolumn{1}{l}{benzophenone }                 & VP  & 1130\,\, & 17     & 308 & 423 & 0.4 & \multicolumn{1}{c}{--}              & 0.008& 3.2  \dots 4.1    & 0.013\dots 0.016 & 3.8  & \multicolumn{1}{r}{\cite{MonteSantos:2006:New-Static_a}}
	\rule[-1.2ex]{0pt}{0pt}\B \\ \hline
	\multicolumn{13}{l}{\textsuperscript{a} VP -- vapour pressure, ACS -- absorption cross section, SCS -- scattering cross section. } \T \\ 
	\multicolumn{13}{l}{\textsuperscript{b} From eq.~(\ref{eq:diam}) at 298\,K using data given in Ref.~\cite{GreenPerry:2008:Perrys_a}. The average temperature has been used to calculate the thermal transpiration correction. } \\   
	\multicolumn{13}{l}{\textsuperscript{c} Corrections following our discussion, we give a range corresponding to variation of the effective diameter by $\pm 11\,\%$ for the Miller equation.} \\  
	\multicolumn{13}{l}{\textsuperscript{d} Original correction based on the TS extrapolation formulae \cite{PoulterRodgers:1983:Thermal,TakaishiSensui:1963:Thermal}; no entry means that the correction is insignificant. }\\
	\multicolumn{13}{l}{\textsuperscript{e} Two transpiration stages with indicated diameters and the temperature sequence 87, 298 and 318\,K. }\\
	\multicolumn{13}{l}{\textsuperscript{f} Reference measurements in order to explore the role of thermal transpiration.}\\
	\end{tabular}
\end{table*}

\begin{figure}
	\begin{center}
		\includegraphics[width=8.5cm]{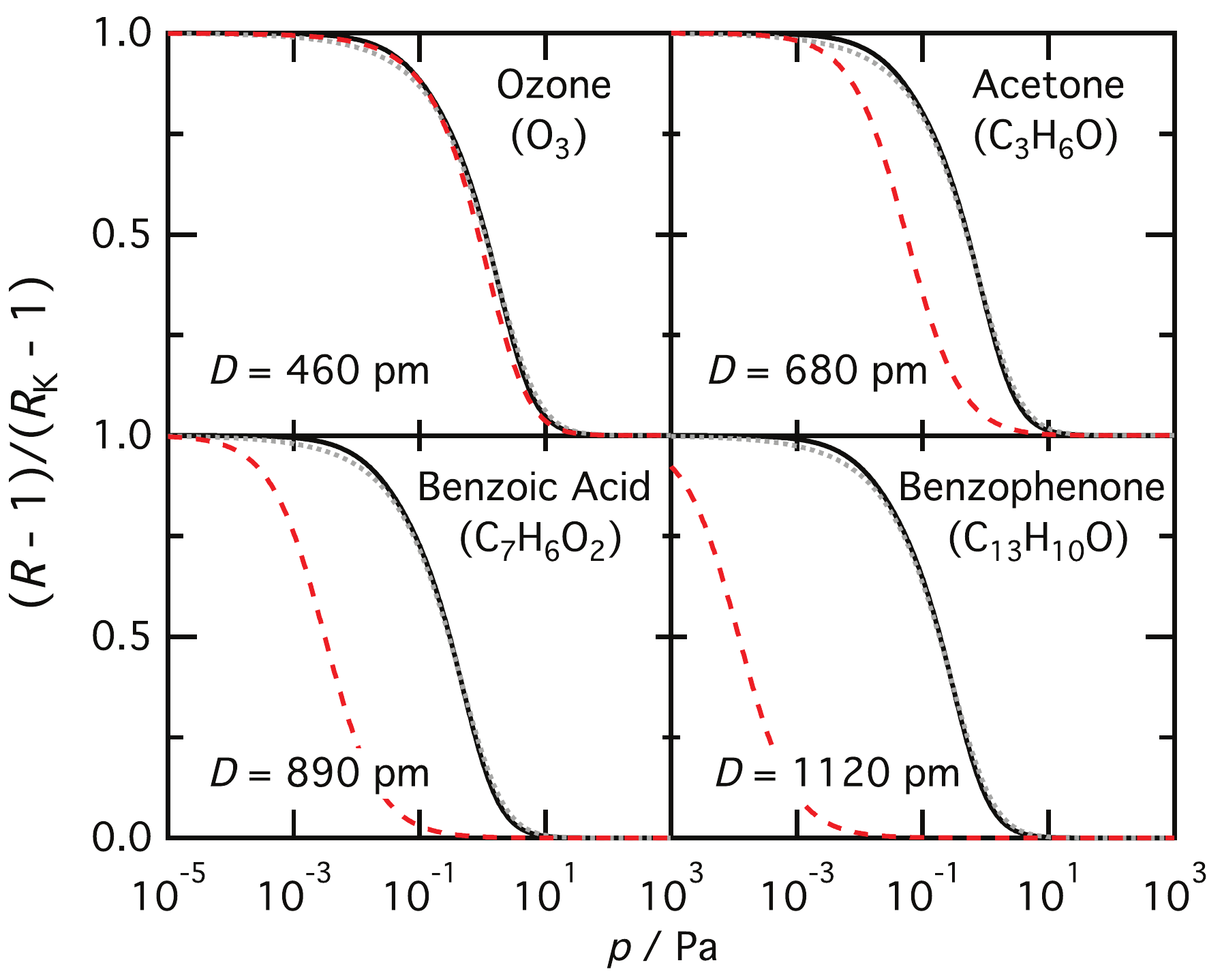} \caption{\label{fig:TranspirationExamples} Molecule dependence of thermal transpiration corrections exemplified by the thermal transpiration curves \(  \left(R-1\right)\left/\left(R_{\mathrm{K}}-1\right)\right. \) of ozone, acetone, benzoic acid and benzophenone. Calculations have been done for $T_1 = 296$\,K, $T_2=318.15$\,K and $d=10$\,mm. Black line -- Miller, red dashed line -- Takaishi \& Sensui (TS), and grey points -- \v{S}etina curve. Due to their transition pressures $p_{1/2}$ being linked directly (eqs.~(\ref{eq:SetHalfP}) and (\ref{eq:MillerHalfP})), the Miller and \v{S}etina models always overlap. Molecular diameters $D$ have been obtained from viscosity data in Ref.~\cite{GreenPerry:2008:Perrys_a}, using eq.~(\ref{eq:diam}) and $T=307\,$K. }  
	\end{center}
\end{figure}

Table~\ref{tab:earlierresults} once more demonstrates that the equations of Miller and of \v{S}etina essentially yield the same corrections and we can use either of the two to compare with those from TS.  Miller corrections are always in the few percent range, sometimes limiting the precision of the measurements. When comparing TS and Miller models, two counterbalancing effects become apparent, that are also illustrated in Figs.~\ref{fig:gasdep} and \ref{fig:TranspirationExamples}. First, the discrepancy between  transition pressures $p_{1/2}$ predicted by TS and the other two models increases with increasing molecular diameter, leading to TS corrections becoming smaller with increasing diameter while Miller and \v{S}etina corrections remain appreciable. However, and albeit weaker this is the second effect,  the transition between molecular and viscous regimes also slightly shifts towards lower pressures with increasing molecular size. Thus on the one hand, predictions by the TS equation are getting worse with increasing diameter, but on the other hand, transpiration effects often become less important in real systems, because relevant pressure scales are more  difficult to reach. Only certain measurements, which require a high degree of precision need to be corrected for thermal transpiration.  The vapour pressure measurements of ozone and the absorption cross section measurements of ethane are such examples and they illustrate that all three models yield similar results for small molecules. The agreement of the different models for ozone becomes also apparent from Fig.~\ref{fig:TranspirationExamples}. The positron scattering experiments of the lighter molecules ethylene,  ethane and 1,4 dioxane also require quasi model independent pressure corrections -- this time because pressures are so low that the model independent low pressure limit is almost 
reached; thus the interpretation of these data will not change. But for benzene or cyclohexane the situation is very different and large differences arise between the TS correction on the one hand and Miller and \v{S}etina equations on the other hand. As expected, discrepancies are largest for the heaviest molecules (Fig.~\ref{fig:TranspirationExamples}) and the corresponding vapour pressure measurements should be corrected accordingly. We also note that the absorption cross section measurements of acetone would need a correction if an uncertainty better than 2\,\% is required.

It appears that the most important consequence of accepting the Miller or \v{S}etina corrections is its impact on precision vapour pressure measurements of large molecules. First, these measurements cover the pressure range  where according to Miller and \v{S}etina the thermal transpiration effect has almost  reached its maximum value, but where the TS model predicts no effect yet ($\lesssim 0.1\,\%$, see also Fig.~\ref{fig:TranspirationExamples}). Second, these measurements also require very low measurement uncertainties such that even small corrections become important. We note that the calculated corrections of a few 10 mPa are close to the precision of the measurements. Their neglect thus constitutes an important bias, if gas surface interactions don't strongly weaken the estimation provided by  Miller's equation. \citet{MonteSantos:2006:New-Static_a} give a measurement uncertainty of only about 10\,mPa for their measurements of naphthalene, benzoic acid and benzophenone at 0.4 Pa. At this level of uncertainty, the corrections for thermal transpiration between 13 and 35\,mPa need certainly to be applied. The recommended vapour pressures of naphthalene by \citet{RuzickaFulem:2005:Recommended_a} are somewhat less affected, because estimated corrections are smaller than the measurement uncertainty of about 50\,mPa. But also from these measurements it becomes clear that further improvements on the measurement uncertainty will require a correction for thermal transpiration.  

 The same authors have already pointed out that there is little experimental evidence on thermal transpiration of larger polyatomic molecules. They therefore sought to determine its impact on their  vapour pressure measurements by changing the head temperature from 
323.15 to 473.15\,K when keeping solid naphthalene at 283.48\,K. No pressure change has been observed, and it was concluded that only the TS equation could correctly reproduce the observation. 
It must be noted however, that the uncertainty $u(p) = 0.05\,\mathrm{Pa}+0.005\,p$ of the pressure measurement at 2.7\,Pa corresponds to an extended relative uncertainty  $u_r\left[p_1(323.15\,\mathrm{K})/p_2(473.15\,\mathrm{K})\right] = 6.7\,\%$ using the coverage factor $k=2$.  At the significance level of 5\,\%, the measurement result is therefore equally compatible with the \v{S}etina or Miller equations (see Table~\ref{tab:earlierresults}).

Given that from a physical point of view  thermal transpiration needs to scale  with the inverse Knudsen number or the rarefaction parameter, which we could confirm by comparing argon and styrene, we are convinced that the proposed approach is reliable even though some modifications due to gas specific interactions with the wall cannot be excluded. Concerning the corrections in Table~\ref{tab:earlierresults}, it should also be noted that both the Miller and the \v{S}etina equation are less applicable to experiments with large $T$ differences than to measurements where the gauge temperature ($T_2$) is close to the measurement temperature ($T_1$), because the original equations have been derived as approximations for small temperature differences.

\section{Conclusions}  

We have established a simple experimental technique to determine thermal transpiration effects in capacitive diaphragm gauges and we have studied the thermal transpiration correction for argon and styrene. 

Using argon as a test gas, three out of numerous semi empirical models have been identified to show best agreement with the  data, confirming an earlier study \cite{YoshidaKomatsu:2010:Compensation_a} on nitrogen. Similar to earlier observations \cite{JitschinRohl:1987:Quantitative}, we also need to introduce an effective diameter, which we tentatively attribute to gas-surface effects. 

Our measurements on styrene (C\( _8 \)H\( _8 \))  demonstrate for the first time that the currently recommended application of the TS correction fails for large molecules, given the non-physical scaling of the transition pressures. The example of styrene shows that the characteristic pressure is underestimated by a factor of 60$-$70, but the degree of underestimation is a growing function of increasing molecular diameter. Quantitative corrections using formulae in eqs.~(\ref{eq:TSCoefA})-(\ref{eq:TSCoefC}) thus are invalidated and studies of large molecules who blindly rely thereon need to be re-checked. 

The recent modification proposed by \citet{Setina:1999:New-approach_a} and the alternative \citet{Miller:1963:On-the-Calculation_a} equation, which scale with $Kn$ through $p_{1/2} \simeq 2 \eta \cdot v_{th} /d$ provide a much better and physically motivated description of the gas dependence of thermal transpiration. Based on the agreement with our Ar measurements, we prefer the use of the Miller equation. However, more studies are required to assess the accuracy of both of the two correction schemes for large diameter molecules. The applicability of the \v{S}etina equation has so far been verified on Ar, N\( _2 \), H\( _2 \) and He  with maximum deviations of 0.1\,\% \cite{Setina:1999:New-approach_a}. Similar investigations of the Miller equation using CDGs are based on  N\( _2 \) \cite{YoshidaKomatsu:2010:Compensation_a}, and Ar (this work), still a non-negligible source of uncertainty being the gas surface interaction.  Both of these studies show that the Miller approach reaches the same or even a better degree of agreement with experiments. The discussion of real world examples shows that  low temperature vapour pressure studies and positron scattering measurements will be affected, due to the low pressures or temperatures employed and that improving the uncertainties will also depend on adequately correcting for thermal transpiration. 

Given that both  theoretical and experimental studies of the thermal transpiration of large polyatomic molecules are sparse and challenging, much remains to be done for developing reliable schemes for accurately predicting thermal transpiration of these species, in particular when gas-surface interactions need to be taken into account. The good agreement between the Miller correction and our measurements of styrene, however, seems to indicate that a simple and general phenomenological approach is possible, at least when temperature differences are not too high. 
   
\appendix 
\section{Thermal transpiration equations} \label{sec:TTE}
The following equations are given for convenience and without derivation.  We refer to the original literature for more details. As before, \( T_1 < T_2 \) are sample and sensor temperatures and \( p_1 \) and \( p_2 \) the corresponding pressures. We keep the convenient definition of the pressure ratio \( R=p_1/p_2 \) and its Knudsen limit \( R_{\mathrm{K}} =\sqrt{T_1/T_2}\). \( d \) is the diameter of the connecting tube and \( \lambda  \) the mean free path length. 
         
\subsection{Liang equation}    
The \citet{Liang:1953:On-the-calculation_a} equation is the most simple of the type of equations (eq.~\ref{eq:thermaltrans}) discussed in the main text with \( f=1 \).
\begin{equation}      \label{eq:Liang}
	     \theta=\frac{1-R}{ 1-R_{\mathrm{K}}} = \left(\alpha x^2+\beta x+1\right)^{-1}\,.
\end{equation}    
Here \( x = \phi_g p_2 d \), \( \beta = 5.76\, \left(1-R_{\mathrm{K}}\right) \)\,(m\,Pa)$^{-1}$ and \(  \alpha =\) \mbox{1.42\,(m\,Pa)$^{-2}$}, where \( \phi_g \) is an empirical gas dependent scaling factor which takes the values \( \phi_{\rm{He}} = 1\) for helium and \( \phi_{\rm{Ar}} = 2.93\) for argon. Note that the gas dependence inherent in \( \phi_g \) scales with the molecular diameter.     

\subsection{Modification of Bennet and Tompkins}    
Based on their measurements and a critical review of the available literature, \citet{BennettTompkins:1957:Thermal_a}  introduced a temperature dependence into the parameter \( \alpha \) of Liangs equation: \( \alpha = 2.08\,  \left(1.70-2.6 \cdot 10^{-3} (T_2-T_1)\right)^2 \)\,Pa\( ^{-2} \)\,m\( ^{-2} \). They also slightly modified the values of  $x$, \( \beta \) and \( \phi_g \):  $x=f \phi_g p_2 d$, where $f=1.22$ for $d>1\,$cm and $f=1$ otherwise, \( \beta = 5.91\, \left(1-\sqrt{T_1/T_2}\right) \)\,Pa\( ^{-1} \)\,m\( ^{-1} \), \( \phi_{\rm He}= 1 \), and \( \phi_{\rm Ar}= 2.70 \). Again, the gas dependent coefficient \( \phi_g \) scales with the molecular diameter.

 \subsection{Kavtaradze equation}    
 
\citet{Kavtaradze:1954:Vliyanie_a} has derived the following expression 
\begin{equation}      \label{eq:Kavtaradze}
	     p_2 d \vartheta = \frac{\ln\left(R/R_{\mathrm{K}}\right)}{1-R},
\end{equation}
where the symbols have their previously defined meanings and \( \vartheta = 1/( p_2 \lambda)\) is calculated using the Sutherland correction for the molecular diameter. For argon, the values \(\vartheta =  \sigma^2_\infty (1 + C / \overline{T}) \), with \( \overline{T} = (T_1 + T_2)/2 \), \( C= 142\)\,K, and  \( \sigma_\infty = 242 \dots 367\,\mathrm{pm} \) have been used. The curve in Fig.~\ref{fig:ThermTransArgon} has been produced using  \( \sigma_\infty = 300 \)\,pm. If, for reasons of consistency, we calculate \( \lambda \) using equations~(\ref{eq:diam}) and (\ref{eq:doverlambda}), this curve is shifted by -5.4\,\% towards lower pressures.

\subsection{KIK equation}   \label{sec:Kanki} 
The Kanki$-$Iuchi and Kosugi (KIK) \cite{KankiIuchi:1976:Flow_a} equation reads 
\begin{equation}      \label{eq:Kanki2}
	  \ln \left( R\right)    = \Omega\left(x \right) \ln \left( R^2_{\mathrm{K}}\right),
\end{equation}                                                                               
where  \( x= d/\lambda \)   and
\begin{equation}      \label{eq:OmegaKanki}
	  \Omega\left(x \right)    = \frac{C^\star}{\frac{\pi}{32}x^2+\frac{9 \pi}{32}x+\frac{4}{3}}
\end{equation} 
with an empirical constant \( C^\star\), which must be equal to \( 2/3 \) if the expression is required to reach the Knudsen limit. 

\subsection{Ebert--Albrand equation}\label{sec:EbertAlbrand}   
 The \citet{EbertAlbrand:1963:The-applicability_a} equation is obtained from integrating eq.~(\ref{eq:diffthermaltrans})  by extending Knudsens low pressure approximation \( \Theta = 1/ ( 1 + d / \lambda ) \) to all the pressure range and assuming that its pressure and temperature dependence can be neglected during integration:
\begin{equation}      \label{eq:EbertAlbrand}
   R    =  R_\mathrm{K}^{\left( 1 + d / \lambda \right)^{-1}}.
\end{equation}                                                                               

No explicit formula is given for the calculation of \( d/\lambda \). In this paper we have made use of eq.~(\ref{eq:doverlambda}) or   (\ref{eq:xmiller}).    
\section*{Acknowledgements}
The authors acknowledge funding from the INSU/CNRS program LEFE-CHAT (MMAIO) and the French ANR IDEO (ANR-09-BLAN-0022-03) and they are grateful to two anonymous reviewers who pointed out flaws in previous versions of the manuscript. The authors are particularly thankful to Denise Mahler-Andersen, who helped with the manipulation of styrene and also to the staff of MKS in France, Germany and the US for providing information on their instruments and calibration. B. D. thanks the French government and UPMC for financing his doctoral work and  C. J. acknowledges fruitful and extended discussions with Peter F. Bernath, Michael J. Brunger, Luca Chiari, Jeremy J. Harrison, Michal Fulem, Kv\v{e}toslav R\r{u}\v{z}i\v{z}ka, and Hajime Yoshida, who openly communicated details on their experiments. 

\end{document}